\documentclass[12pt,a4paper]{article}

\usepackage{amsfonts}
\usepackage{amssymb,amsmath}
\usepackage{graphicx}

\newcommand{\be}{\begin{equation}}
\newcommand{\ee}{\end{equation}}
\newcommand{\bea}{\begin{eqnarray}}
\newcommand{\eea}{\end{eqnarray}}
\newcommand{\la}{\langle}
\newcommand{\ra}{\rangle}

\newcommand{\ld}{\left(}
\newcommand{\rd}{\right)}
\newcommand{\lb}{\left\{}
\newcommand{\rb}{\right\}}
\newcommand{\lbr}{\left[}
\newcommand{\rbr}{\right]}
\newcommand{\rabg}{\rangle_{b,\,g}}
\newcommand{\rabgp}{\rangle_{b,\,g'}}
\newcommand{\rapg}{\rangle_{p,\,g}}
\newcommand{\rapgp}{\rangle_{p,\,g'}}
\newcommand{\non}{\nonumber \\ }
\newcommand{\ve}{\varepsilon}

\begin{document}

\title{Ground-state correlations of itinerant electrons\\
       in the spinless Falicov-Kimball chain\\
        and related tight-binding systems}

\author{Janusz J\c{e}drzejewski$^{1,2,\star}$,
        Taras Krokhmalskii$^{3,1,\ast}$
        and
        Oleg Derzhko$^{3,4,\dagger}$\\
\sf        $^{1}$Institute of Theoretical Physics,
                    University of Wroc\l aw\\
\sf        Max Born Sq. 9, 50-204 Wroc\l aw, Poland\\
\sf        $^{2}$Department of Theoretical Physics,
                    University of \L \'{o}d\'{z}\\
\sf        149/153 Pomorska Str., 90-236 \L \'{o}d\'{z}, Poland\\
\sf        $^{3}$Institute for Condensed Matter Physics\\
\sf        1 Svientsitskii Str., L'viv-11, 79011, Ukraine\\
\sf        $^{4}$Chair of Theoretical Physics,
                    Ivan Franko National University of L'viv\\
\sf        12 Drahomanov Str., L'viv-5, 79005, Ukraine}

\date{\today}

\maketitle

\vspace{-8mm}
\begin{abstract}

We consider the one-dimensional spinless Falicov-Kimball model of
itinerant fermionic particles (``spinless electrons''), which can
hop between nearest-neighbour sites only, and of immobile
particles (``classical ions''), with an on-site attraction. Extensive
studies of the ground-state phase diagram of this system and its
higher dimensional counterparts, carried out up to now,
concentrated on determining ground-state arrangements of ions on
the underlying lattice, while the properties of electrons were
typically ignored. We report studies of short- and long-range
correlations between electrons, and between ions and electrons,
and of the spatial decay of electron correlations
(decay of single-particle density matrix), in the ground state.
The studies have been carried out analytically and by means of
well-controlled numerical procedures.
In the case of period 2 ground state, the single-particle density matrix
has been expressed in terms of a hypergeometric function,
and its spatial decay has been extracted.
Numerical calculations have been done for open chains of various
lengths (up to a few thousand sites), in order to control the
chain-size dependence of correlations and to extrapolate the results to
the limit of infinite chain.
A part of the obtained results refers to tight-binding electrons
subjected to a periodic external potential due to the ions, which
constitute simple models of metals and insulators.
\end{abstract}

\vspace{1mm}

\vspace{1mm}


\vspace{5mm}


\noindent
E-mail addresses: \\
$\star$ --- jjed@ift.uni.wroc.pl \\
$\ast$ --- krokhm@icmp.lviv.ua\\
$\dagger$ --- derzhko@icmp.lviv.ua\\

\newpage
\section{Introduction}

The Falicov-Kimball model emerged from solid-state theory in 1969,
as a simple model for semiconductor-metal transitions \cite{falkim}.
Since that time it has become an important standard tool
for studying various properties and phenomena
that occur in strongly correlated fermionic systems,
ranging from metal-insulator and mixed-valence phenomena
to the Peierls instabilities and crystallization.
For an overview and an extensive list of relevant references
the reader is advised to consult \cite{grumac,jele}.

In the present paper we are studying the simplest, skeleton
version of the model proposed by Falicov and Kimball, which is
usually referred to as the spinless Falicov-Kimball model. This
is a lattice system that consists of two sorts of spinless
fermions, called electrons and ions. The electrons can hop
between nearest-neighbour sites only, while the ions are
immobile. The only interaction occurs between the electrons and
ions and it is an on-site attraction of strength $U$. Throughout
this paper, the underlying lattice is one-dimensional (later on
called the chain), so it can be identified with integers $\mathbb{Z}$.
When the chain is finite and consists of $L$ sites labeled $x=0,
\ldots , L-1$, we impose {\em boundary conditions} $b$, which
in this paper are either free ($b=f$) or periodic ($b=p$). Then,
the whole two-component system is governed by the Hamiltonian:

\begin{equation}
\label{H}
H_b= - \sum_{x=0}^{L-1} \left (t_{b,x} a^+_x a_{x+1}+
t_{b,x} a^+_{x+1}a_x + U w_x a^+_x a_x\right),
\end{equation}
where $a^+_x$, $a_x$ are the creation and annihilation operators
of an electron at the site $x$ of the chain, $t_{b,x}$ is the
hopping rate between the sites $x$ and $x+1$ (which will be set
site-independent, except at the boundary of a chain), and $w_x$
stands for the occupation-number operator of an ion at site $x$.
Since the occupation-number operators $w_{x}$ commute with the
Hamiltonian (\ref{H}), we can identify them with classical
variables taking values zero or one. Consequently the total
number of ions, $N_i=\sum_{x}w_x$, is conserved. Moreover, the
total particle-number operator of electrons, $\hat{N}_e
=\sum_{x}a_x^+ a_x$, commutes with $H_b$, so the total number of
electrons, $N_{e}$, is also conserved. Therefore, we can study the
equilibrium properties of our system in the canonical ensemble,
with a fixed number of sites $L$, the electron number $N_e$ and the
ion number $N_i$. The corresponding canonical partition function
reads:
\begin{equation}
\label{parf}
Z_b= \sum_{w: \sum_{x} w_{x} = N_{i}}{\mbox {Tr}}_{N_{e}} \exp \left(-\beta
H_{b} \right),
\end{equation}
where $w$ stands for the function that to a site $x$ of the
underlying chain assigns the value $w_x=0,1$, which can be viewed
as a sequence $w=w_0 w_1 \ldots$ (hereafter called the {\em ion
configuration}), the trace is over the Hilbert space of $N_e$
electrons, and $\beta$ is the inverse temperature.

Despite the apparent simplicity of the form of Hamiltonian
(\ref{H}) and its counterparts on higher-dimensional lattices
$\mathbb{Z}^d$, $d>1$,
the question
whether the system is capable of describing cooperative effects
had remained open for almost two decades. It was affirmatively
answered in 1986 \cite{brasch,kennlieb}, by demonstrating that
the half-filled system exhibits a staggered type of long-range
order in the ionic subsystem at zero temperature, and showing
that this staggered order persists at sufficiently low
temperatures in two- and higher-dimensional systems
\cite{kennlieb}.

Since then, one can observe an increased interest in the model
and many properties of the Falicov-Kimball model have been
established by means of rigorous methods as well as by means of
various sorts of numerical procedures. Not surprisingly, a vast
majority of obtained results refers to the ground state (see
\cite{grumac,gruber,jele,flu} and references quoted there). As
there are good reasons to believe that studies of the ground state
of the one-dimensional model may reveal interesting cooperative
phenomena, which herald the appearance of similar cooperative
phenomena in higher-dimensional systems at zero and sufficiently
low temperatures, a lot of efforts at determining ground-state
properties have been made
\cite{freefal}--\cite{gale}.

As a matter of fact, all the attention, not only in the last
quoted papers but also in many others whose references can be
found in \cite{grumac,gruber,jele,flu}, has been concentrated on
the ionic subsystem. Specifically, the most favourable
energetically ionic configurations and ionic correlations, at
different values of the electron-ion coupling parameter $U$, have
been determined. The problem of correlations between the
itinerant fermionic particles had not been usually addressed. To
the best of our knowledge, there are just a few exceptions. As
mentioned in \cite{grumac} the results on decay of
superconducting correlations, obtained in \cite{komatas,maruiz}
for general itinerant-electron systems, can be
applied to the ground state of the one-dimensional
Falicov-Kimball model (\ref{H}). In particular, it follows that
the ground-state average of $a^+_xa_y$ decays exponentially with the
distance between the sites $x$ and $y$, i.e. there is no so called
off-diagonal long-range order. Similar results, but for
multi-point products of fermionic creation and annihilation
operators (that cannot be expressed in terms of occupation-number
operators), have been obtained in the strong coupling regime in
\cite{messager1}. In \cite{lebmac} a class of models related with the
spinless Falicov-Kimball model, with an emphasis on the static
Holstein model, has been analyzed rigorously. As a byproduct of
considerations carried out for the static Holstein model, the
authors have obtained a relation between grand-canonical averages
of one-site ion- and one-site electron-occupation numbers at
half-filling. This result seems to be the first one referring to
correlations between the electrons that can be expressed as an average
of electron occupation-number operators.
The mentioned relation implies that, at half-filling, the staggered
long-range order in the ion subsystem is accompanied by the same
kind of long-range order in the electron subsystem.

It should be emphasized here that we are concerned only with
low-dimensional systems. There is an extensive literature
concerning the limit of infinite dimensions, where calculations
of some correlation functions, like linear-response functions, are quite
common (see for instance \cite{freezla}).

The purpose of this paper is to examine the properties of the
electron subsystem of the model (\ref{H}), at zero temperature.
It is achieved by means of a number of short- and long-range
correlation functions of electrons, and electrons and ions, that
can be represented as canonical averages of operators constructed
out of occupation-number operators of electrons and ions.

The paper is organized as follows.
We start, in Section 2, with a brief discussion of
canonical ground-state averages. In Section 3, we define
and discuss some general properties of short- and long-range
correlation functions, to be studied in the subsequent sections.
Then, in Section 4,
we present analytical results for closed finite chains and
for infinite chains.
After that, in Section 5, we give numerical results, obtained by means
of numerical exact diagonalization.
We end up, in Section 6, with a discussion and a summary of the obtained
results.

\section{The canonical ground state and ground-state averages}

There are two approaches that enable one to study ground-state
properties of a system. The first one is a quantum-mechanical
approach, usually adopted in the papers concerning ground-state of
Falicov-Kimball models, and consists in considering a finite (and
then an infinite) system exclusively at zero temperature. The
second one is a thermodynamic approach, where the considerations
start with nonzero temperatures and then the limit of zero
temperature is taken.

In the quantum-mechanical approach, one makes use of the fact,
mentioned in Introduction, that a system of $N_e$ electrons and
$N_i$ ions with Hamiltonian (\ref{H}), confined to a chain (of
$L$ sites and with boundary conditions $b$), can be considered
for a fixed ion configuration $w$.
The corresponding Hamiltonian, denoted $H_b (w)$,
describes tight-binding electrons in the external
potential $-Uw_x$. In this case,
the ground state of $N_e$ electrons, $|w\ra_{b}$,
that is the eigenvector of $H_b (w)$ to the lowest
energy $E_b (w)$, is the so called {\em Fermi-sea state}. Then,
the ground state of a finite chain with
$N_e$ electrons and $N_i$ ions is given by the set $G_b$ of
{\em finite-chain ground-state ion configurations} $g_b$ and the
corresponding Fermi-sea states $|g_{b}\ra_{b}$ of energy $E_b (g_b)$,
such that
\begin{equation}
\label{Eb}
E_b (g_{b})= \min_{w: \sum_{x} w_{x} = N_{i}} E_b (w).
\end{equation}
To get thermodynamically relevant quantities, we first introduce
the finite-chain ground-state energy densities:
\be
\label{eb}
e_b (w) = L^{-1} E_b (w), \  \
e_b (g_b) = L^{-1} E_b (g_b) = \min_{w: \sum_{x} w_{x} = N_{i}} e_b (w),
\ee
for a given ion configuration $w$,
and for given particle numbers $N_e$, $N_i$, respectively.
Then, we go to the thermodynamic limit,
denoted $\lim_{L \rightarrow  \infty}$, i.e.
we send $L$ to infinity in such a way that
the both ends of the chain (if any) become remote from any fixed site,
and
$L^{-1}N_e \rightarrow \rho_e$ -- the electron density and
$L^{-1}N_i \rightarrow \rho_i$ -- the ion density,
for some $\rho_e, \rho_i \in [0,1]$.
In the thermodynamic limit and for a fixed ion (infinite-chain)
configuration $w$, the  (infinite-chain)
ground-state energy density is
\be
\label{ew}
e(w)= \lim_{L \rightarrow \infty} e_b (w) ,
\ee
and is independent of boundary conditions. In the r.h.s of (\ref{ew}),
the ion configuration $w$ stands for
the restriction of the infinite-chain configuration $w$ to the
finite chain. For given particle densities
$\rho_e$, $\rho_i$, the {\em infinite-chain ground-state energy
density}, $e$, is
\begin{equation}
\label{eew}
e= \min_{w} e(w) = e(g),\ \  g \in G,
\end{equation}
where the minimum is taken over infinite-chain configurations of
ions and the set $G$, defined by (\ref{eew}), stands for the set of
{\em infinite-chain ground-state ion configurations}.
Alternatively, the ground-state energy density can be obtained as
\be
\label{eeb}
e= \lim_{L \rightarrow \infty} e_b (g_b), \  \   g_b \in G_b.
\ee

In the thermodynamic approach, we start
with nonzero temperatures and define the finite-chain internal-energy density
$e_{\beta, b}$,
\begin{equation}
\label{ebetab}
e_{\beta ,b}= L^{-1} \la H_b \ra_{\beta, b},
\end{equation}
where
\begin{equation}
\label{canavH}
\la H_b \ra_{\beta ,b} = Z_{b}^{-1} \sum_{w: \sum_{x} w_{x} = N_{i}}
{\mbox {Tr}}_{N_e} H_b \exp(-\beta H_b)
\end{equation}
is the canonical average of $H_b$.
Then, the (infinite-chain) internal-energy density,
$e_{\beta}$, is given by
$e_{\beta} = \lim_{L \rightarrow \infty} e_{\beta ,b}$,
and is independent of boundary conditions.
Thermodynamically, the ground-state energy density, $e$,
equals to $e= \lim_{\beta \rightarrow \infty} e_{\beta}$.
Reversing the order of the limits $L \rightarrow \infty$,
$\beta \rightarrow \infty$,
we recover the formula (\ref{eeb}) of the quantum-mechanical approach
\be
\label{eebeta}
e = \lim_{L \rightarrow \infty } e_b (g_b), \  \
e_b (g_b) = \lim_{\beta \rightarrow \infty} e_{\beta ,b}=
L^{-1} {_{b} \la } g_{b}| H_b | g_{b} \ra_{b}, \  \  g_b \in G_b.
\ee
To define a ground-state average, we start with nonzero
temperatures. Let $A$ be an operator built out of $a^+_x$, $a_x$ and/or
$w_x$, depending on a finite number of sites only. By definition, the
finite-chain canonical average of $A$, $\la A \ra_{\beta , b}$,
is
\be
\label{canavb}
\la A \ra_{\beta , b} = Z_{b}^{-1} \sum_{w: \sum_{x} w_{x} = N_{i}}
{\mbox {Tr}}_{N_e} A \exp(-\beta H_b).
\ee
The infinite-chain limit of $\la A \ra_{\beta , b}$ amounts to the canonical
average of $A$, $\la A \ra_{\beta}$,
\be
\label{canav}
\lim_{L \rightarrow \infty} \la A \ra_{\beta , b} = \la A \ra_{\beta},
\ee
which is independent of boundary conditions and translation invariant.
Then, the {\em infinite-chain ground-state average} of $A$, $\la A \ra$,
is defined as
\be
\label{gsav_beta}
\lim_{\beta \rightarrow \infty} \la A \ra_{\beta} = \la A \ra .
\ee
The ground-state average $\la A \ra$ shares the above mentioned properties
of $\la A \ra_{\beta}$.

By rigorous results of \cite{lemberger} and numerical results of
\cite{guj, gjl,farky}, for pairs of rational densities $(\rho_e,
\rho_i)$, satisfying a linear relation, and for suitable values
of electron-ion attraction $U$, the set $G$ of ground-state
configurations of ions consists of a finite number of elements.
This happens, for instance, if $\rho_e=p/q$, where the natural
numbers $p$ and $q$ are relatively prime, and $\rho_i = \rho_e$
(the so called neutral case), for sufficiently large $U$ \cite{guj,gjl}.
Then, $G$ consists of $q$ periodic configurations of period $q$ (the so
called most homogeneous configurations) that differ from each
other by a translation. In such a case, the ground-state average
of $A$ equals to
\be
\label{gsav}
\la A \ra = q^{-1} \sum_{g \in G} \lim_{L \rightarrow \infty}
{_{b} \la } g | A | g \ra_{b} ,
\ee
where $g$ in $| g \ra_b$ stands for the restriction of a $g \in G$
to the finite chain. As a finite-chain approximation to $\la A
\ra$, converging to $\la A \ra$ as $L \rightarrow \infty$, we
choose $\la A \ra_b$  given by
\be
\label{gsavb}
\la A \ra_b =
q^{-1}\sum_{g \in G} {_{b} \la } g | A | g \ra_{b} .
\ee
In what follows, we refer to the ground-state average  (\ref{gsavb})
as the {\em {symmetric average}}, while the quantum average
${_{b} \la } g | A | g \ra_{b}$ is called
the {\em {broken-symmetry average}}.
Since the quantum averages ${_{b} \la } g | A | g \ra_{b}$
appear frequently throughout the paper, we introduce the
abbreviated notation
$\la A \ra_{b,g} \equiv {_{b} \la } g | A | g \ra_{b}$.
In general, a finite set of $Q$ ground-state
configurations $G$ may consist of configurations related by
translations and reflections, with $Q\geq q$.

\section{Short- and long-range correlation functions: definitions,
general properties, and the method of calculation}

Throughout the sections that follow, the set $G$ stands for the
set of periodic ground-state configurations of the infinite chain,
for given $\rho_e=p/q$, with $p$ and $q$ being relatively prime
natural numbers, $\rho_i = m \rho_e$, with $m=1,2, \ldots$,  and for
suitable values of $U$.
Later on, the set of $U$ for which $G$ is the set of the ground-state
configurations will be called the {\em stability interval} of $G$,
and denoted $\Delta U_G$.
The set $G$ consists of $q$ ion configurations,
with period $q$,
which are known as the atomic $(m=1)$ or $m$-{\em molecular} $(m \geq 2)$
{\em most homogeneous ion configurations} \cite{lemberger,guj,gjl}.
Following \cite{gjl}, the set of
$m$-molecule most homogeneous configurations
corresponding to the electron density $\rho_e = p/q$ is denoted $[p/q]_m$.
The elements of $[p/q]_m$ can conveniently be described by specifying
their restriction to the unit cell, whose sites are labeled
$x=0, 1, \ldots ,q-1$,
i.e. as a sequence $g=g_0 g_1 \ldots g_{q-1}$.
The positions of ions in the unit cell of the {\em representative configuration}
of $[p/q]_m$ can be obtained as the solutions $k_j$ to the equations:
\bea
\label{circlemap}
pk_j = j \ \  \mbox{mod} \ \ q, \ \ \ j=0,1, \ldots, mp-1.
\eea
The remaining $q-1$ configurations of $[p/q]_m$ are obtained by translating
the defined above representative configuration.

Moreover, the number of sites in the chain, $L$,
is set to be a multiple of $q$, so that
$L^{-1}N_e=\rho_e$, and $L^{-1}N_i=\rho_i$.
For the ion subsystem the correlation functions are defined in Appendix A.
Below, we define analogous correlation functions for the electron subsystem.
The list of correlation functions, studied in the sections that follow,
opens a one-point function, the {\em {local electron density
in the ion configuration $g$}}:
\be
\label{nxbg}
\la n_x \ra_{b,\,g}, \ \ \ g \in G.
\ee
It is related to $\rho_e$ by
\bea
\label{nxbgrho}
L^{-1}\sum_x\la n_x \ra_{b,\,g}=\rho_e.
\eea
For periodic boundary conditions ($b=p$), $\la n_x \ra_{p,\,g}$
has period $q$: $\la n_x \ra_{p,\,g}=\la n_{x+q} \ra_{p,\,g}$,
and
\bea
\label{nxp}
\rho_e =\la n_x \ra_p  =
q^{-1} \sum_{g \in G} \la n_x \ra_{p,\,g}=
q^{-1} \sum_{y=0}^{q-1} \la n_{x+y} \ra_{p,\,g} .
\eea
Then, we consider a number of two-point correlation functions.
The first one is the
{\em {density-density correlation in the ion configuration $g$}},
$\la n_xn_y \ra_{b,\,g}$, which (by the Wick's theorem) can be
expressed as
\bea
\label{Wick1}
\la n_xn_y \ra_{b,\,g}=\la n_x \ra_{b,\,g}\la n_y
\ra_{b,\,g}-|\la a^+_xa_{y}\ra_{b,\,g}|^2 .
\eea
For the periodic boundary conditions, $\la n_xn_y \ra_{p,\,g}$
has period $q$, $\la n_xn_y \ra_{p}$ is translation invariant,
and the symmetric average
$\la \ld n_x-\rho_e \rd \ld n_y-\rho_e \rd \ra_p$
can be written as follows
\bea
\label{Wick2}
\la \ld n_x-\rho_e \rd \ld n_y-\rho_e \rd \ra_p=q^{-1}\sum_{g
\in G} \la n_xn_y \ra_{p,\,g} - \rho_e^2  \non
=q^{-1}\sum_{g\in G}
\la n_x - \rho_e \ra_{p,\,g} \la n_y - \rho_e \ra_{p,\,g} -
q^{-1}\sum_{g}|\la a^+_xa_y \ra_{p,\,g}|^2.
\eea
The symmetric average of the above kind is used to define the
{\em electron-electron short-range correlation function},
${\cal{L}}_{b,\,x}(l)$,
\bea
\label{Lbx}
{\cal{L}}_{b,\,x}(l)= \la \ld n_x-\rho_e \rd \ld n_{x+l}-\rho_e \rd \ra_b .
\eea
In an analogous manner, we define the
{\em electron-ion short-range correlation function}, ${\cal{S}}_{b,\,x}(l)$:
\bea
\label{Sbx}
{\cal{S}}_{b,\,x}(l)=q^{-1}\sum_{g\in G}\la \ld g_x-mn_x \rd \ld
g_{x+l}-mn_{x+l} \rd \ra_{b,\,g} .
\eea
In the case of the periodic boundary conditions,
the correlation functions,
${\cal{L}}_{b,\,x}(l)$  and ${\cal{S}}_{b,\,x}(l)$,
are translation invariant (i.e. depend only on $l$).
To define the long-range correlations corresponding to the above
short-range correlations, we start with the
{\em {order-parameter operators}}
$\hat{O}_k$, $\hat{O}_g$, and $\hat{O}_{g,\,g'}$:
\bea
\label{Ok}
\hat{O}_k=L^{-1} \sum_{x=0}^{L-1} \mbox{e}^{ikx}\ld n_x- \rho_e \rd ,
\eea
\bea
\label{Og}
\hat{O}_g=L^{-1} \sum_{x=0}^{L-1} \ld g_x- \rho_i \rd \ld n_x- \rho_e \rd ,
\eea
and
\bea
\label{Ogg}
\hat{O}_{g,\,g'}=L^{-1}\sum_{x=0}^{L-1} \ld g_x-\rho_i \rd
\ld g'_x-mn_x \rd .
\eea
Taking broken-symmetry averages of the order-parameter operators,
we obtain the corresponding {\em {order parameters}}:
\bea
\label{Okbg}
\la \hat{O}_k \rabg =L^{-1}\sum_{x=0}^{L-1}\mbox{e}^{ikx}\la n_x
-\rho_e \rabg,
\eea
\bea
\label{Ogbg}
\la \hat{O}_g \rabgp =L^{-1}\sum_{x=0}^{L-1} \ld g_x-\rho_i \rd
\la n_x-\rho_e \rabgp.
\eea
and
\bea
\label{Oggpbg}
\la \hat{O}_{g,\,g'}\rabgp =L^{-1}\sum_{x=0}^{L-1}\ld g_x-\rho_i
\rd \la g'_x-mn_x \rabgp .
\eea
The symmetric averages of the order-parameter operators vanish:
\bea
\label{symavorpam}
\la \hat{O}_k \ra_b  = \la \hat{O}_g \ra_b =
q^{-1}\sum_{g'\in G} \la \hat{O}_{g,\,g'}\rabgp
\equiv 0 .
\eea
By means of the order-parameter operators we define the
{\em long-range correlation functions}
${\cal{P}}_b(k)$, ${\cal{L}}_b$, and ${\cal{S}}_b$:
\bea
\label{Pb}
{\cal{P}}_b(k)\equiv \la \hat{O}^+_k\hat{O}_k \ra_b \non
=L^{-2}\sum_{x,\,y=0}^{L-1} \mbox{e}^{ik(x-y)} \la \ld n_x -\rho_e \rd
\ld n_y -\rho_e \rd \ra_b,
\eea
\bea
\label{Lb}
{\cal{L}}_b \equiv q^{-1}\sum_{g\in G} \la \hat{O}_g^2 \ra_b \non
=L^{-2}\sum_{x,\,y =0}^{L-1} q^{-1}\sum_{g\in G}
\lbr  \ld g_x-\rho_i \rd \ld g_y-\rho_i \rd  \rbr
\la   \ld n_x-\rho_e\rd \ld n_y-\rho_e\rd    \ra_b ,
\eea
and
\bea
\label{Sb}
{\cal{S}}_b \equiv
q^{-2}\sum_{g,g'\in G}
\la \hat{O}^2_{g,\,g'}\rabgp \non
=L^{-2}\sum_{x,\,y=0}^{L-1}
q^{-1}\sum_{g\in G}
\lbr \ld g_x-\rho_i \rd  \ld g_y-\rho_i \rd \rbr
q^{-1}\sum_{g'\in G}
\la \ld g'_x-mn_x \rd  \ld g'_y-mn_y \rd \rabgp .
\eea
In the case of the periodic boundary conditions, the long-range correlation
functions ${\cal{P}}_p(k)$, ${\cal{L}}_p $, and ${\cal{S}}_p $ assume the form
\bea
\label{Pp}
{\cal{P}}_p(k) =  L^{-1} \sum_{l=0}^{L-1}
\mbox{e}^{-ikl} {\cal{L}}_p(l),
\eea
\bea
\label{Lp}
{\cal{L}}_p = L^{-1} \sum_{l=0}^{L-1}
E_p(l) {\cal{L}}_p(l),
\eea
\bea
\label{Sp}
{\cal{S}}_p = L^{-1} \sum_{l=0}^{L-1}
E_p(l) {\cal{S}}_p(l) ,
\eea
where ${\cal{L}}_p(l)\equiv {\cal{L}}_{p,\,x}(l)$,
${\cal{S}}_p(l)\equiv {\cal{S}}_{p,\,x}(l)$, and $E_p(l)$ is defined
in Appendix A.
The function $k \rightarrow {\cal{P}}_b(k)$ is usually referred to
as the {\em static structure factor} \cite{mahan}.
Since, the following relations hold,
\bea
\label{Pbg}
\la \hat{O}^+_k\hat{O}_k \rabg =
L^{-2}\sum_{x,\,y=0}^{L-1}
\mbox{e}^{ik(x-y)} \la n_x -\rho_e \rabg \la
n_y - \rho_e \rabg
\non - L^{-2}\sum_{x,\,y=0}^{L-1}
\mbox{e}^{ik(x-y)} |\la a^+_xa_y \rabg|^2 ,
\eea
\bea
\label{Lbg}
\la \hat{O}_g^2\rabgp =
L^{-2}\sum_{x,\,y=0}^{L-1}
\ld  g_x - \rho_i \rd \ld g_y - \rho_i    \rd
\la  n_x - \rho_e \rabgp \la n_y - \rho_e \rabgp
\non -L^{-2}\sum_{x,\,y=0}^{L-1}
\ld  g_x - \rho_i \rd \ld g_y - \rho_i \rd  |\la a_x^+a_y\rabgp|^2 ,
\eea
and
\bea
\label{Sbg}
\la \hat{O}^2_{g,\,g'}\rabgp =
L^{-2}\sum_{x,\,y=0}^{L-1}
\ld g_x - \rho_i \rd  \ld g_y-\rho_i \rd
\la  g'_x - mn_x \rabgp \la g'_y - mn_y  \rabgp
\non -L^{-2}\sum_{x,\,y=0}^{L-1}
\ld  g_x - \rho_i \rd \ld g_y - \rho_i \rd  |\la a_x^+a_y\rabgp|^2 ,
\eea
and since, for periodic boundary conditions, the averages
$\la \hat{O}_g^2 \ra_p$,
$ q^{-1}\sum_{g'\in G} \la \hat{O}^2_{g,\,g'}\rapgp$, and
$\la \hat{O}^+_k \hat{O}_k \rapg$
are translation invariant, hence independent of $g \in G$,
\bea
\label{Pporpar}
{\cal{P}}_p (k) \equiv
\la \hat{O}^+_k\hat{O}_k \ra_p = |\la \hat{O}_k \rapg|^2
- L^{-1}\sum_{l=0}^{L-1}
\mbox{e}^{-ikl} |\la a^+_xa_{x+l} \rapg|^2 ,
\eea
\bea
\label{Lporpar}
{\cal{L}}_p \equiv
q^{-1}\sum_{g\in G} \la \hat{O}_g^2\ra_p =
q^{-1}\sum_{g'\in G}
\ld \la \hat{O}_g \rapgp \rd^2
\non - L^{-1}\sum_{l=0}^{L-1} \ld g_x-\rho_i \rd \ld g_{x+l}-\rho_i \rd
q^{-1}\sum_{g'\in G}| \la a^+_x a_{x+l}\rapgp |^2 ,
\eea
and
\bea
\label{Sporpar}
{\cal{S}}_p \equiv
q^{-2}\sum_{g,g'\in G} \la \hat{O}^2_{g,\,g'}\rapgp =
q^{-1}\sum_{g'\in G} \la \hat{O}^2_{g,\,g'}\rapgp =
q^{-1}\sum_{g'\in G} \ld \la \hat{O}_{g,\,g'}\rapgp \rd^2
\non - L^{-1}\sum_{l=0}^{L-1} \ld g_x-\rho_i \rd \ld g_{x+l}-\rho_i \rd
q^{-1}\sum_{g'\in G}| \la a^+_x a_{x+l}\rapgp |^2 .
\eea
For the one-dimensional systems of the kind considered here,
and in the limit ${L\to \infty}$, the second term in
(\ref{Pporpar}), (\ref{Lporpar}), (\ref{Sporpar}) vanishes
(see \cite{komatas, maruiz} and the sections that follow).
Thus, in the thermodynamic limit we find simple relations between
long-range correlations and the corresponding order parameters
\bea
\label{P}
{\cal{P}}(k) \equiv \lim_{L\to \infty} {\cal {P}}_p (k) =
\lim_{L\to \infty} \la \hat{O}^+_k\hat{O}_k \ra_p
=|\lim_{L\to \infty} \la \hat{O}_k \rapg|^2 ,
\eea
\bea
\label{L}
{\cal{L}} \equiv \lim_{L\to \infty}  {\cal {L}}_p =
q^{-1}\sum_{g'\in G}
\ld \lim_{L\to \infty} \la \hat{O}_g \rapgp \rd^2 ,
\eea
and
\bea
\label{S}
{\cal{S}} \equiv \lim_{L\to \infty} {\cal {S}}_p =
q^{-1}\sum_{g'\in G} \ld \lim_{L\to \infty} \la \hat{O}_{g,\,g'}\rapgp \rd^2 .
\eea
The relations (\ref{P}), (\ref{L}), (\ref{S}) hold for other boundary
conditions as well. If the quantities ${\cal{P}}(k), {\cal{L}}$
are strictly positive, one says that the considered system exhibits
the {\em long-range order}.

In order to calculate the correlation functions defined above, one
can use the fact that for a fixed ion configuration  $w$ the
Hamiltonian of the system, $H_b(w)$, is a second quantized form of
the one-electron Hamiltonian $h_b(w)$. If we denote by
$\{|x\ra \}_{x=0,\ldots,L-1}$ the orthogonal basis of one-electron
states, such that $a^+_x$ creates an electron in the state $|x\ra$,
then the matrix elements of $h_b(w)$ in the basis
$\{ |x \ra \}_{x=0,\ldots,L-1}$ are defined by
\bea
\label{h}
H_b(w)=\sum_{x,\,y=0}^{L-1} \la x|h_b(w)|y \ra a^+_xa_y,
\eea
i.e. explicitly, all the non-vanishing matrix elements are given by
\bea
\label{hxy}
\la x |h_b(w) |x\ra =-Uw_x,\ \ \ \ \ \ \ \la x |h_b(w) |y\ra =-t_{b,\,x} \ \ \ \
\mbox{if} \ \ \ y=x\pm1.
\eea
Let $\{|v\ra\}_{v=v_0,\ldots,v_{L-1}}$ be the orthonormal basis built
out of the eigenstates of $h_b(w)$ to the eigenvalues
$\Lambda_v$, such that $\Lambda_v \leq \Lambda_{v'}$ if $v < v'$.
Then, the unitary matrix $\cal{U}$, with the following matrix elements
${\cal{U}}_{xv}$:
\bea
\label{U}
{\cal{U}}_{xv}=\la x|v\ra,
\eea
diagonalizes the matrix of $h_b(w)$,
\bea
\label{UhU}
\sum_{x,y} {\cal{U}}^+_{vy} \la y|h_b(w) |x \ra {\cal{U}}_{xv'}=
\Lambda_v\delta_{vv'}.
\eea
Moreover, the set of operators $\{b^+_v, b_v\}, \ v=v_0,\ldots,v_{L-1}$,
defined by
\bea
\label{bv}
b_v=\sum_{x=0}^{L-1}\la v|x\ra a_x ,
\eea
satisfies the canonical anticommutation relations, and
\bea
\label{Hbw}
H_b(w)=\sum_{x,\,y=0}^{L-1} \la x|h_b(w)|y\ra
a^+_xa_y=\sum_v\Lambda_vb^+_vb_v.
\eea
Since,
\bea
\label{ax}
a_x=\sum_v \la x|v\ra b_v,
\eea
we can express the basic electron-correlation functions,
$\la n_x\ra$,  $\la a^+_x a_y\ra_{b,\,w}$, and $\la n_xn_y\ra_{b,\,w}$,
in terms of the site-components $\la x|v\ra$ of the eigenvectors $|v\ra$:
\bea
\label{nxny}
\la n_x\ra_{b,\,w} \equiv
{_b\la w|a^+_xa_x|w\ra_b}=
\sum_{v\leq v_F}|\la x|v\ra|^2, \non
\la a^+_xa_y\ra_{b,\,w} \equiv
{_b\la w|a^+_xa_y|w\ra_b}=
\sum_{v\leq v_F}\la v|x\ra \la y|v\ra, \non
\la n_xn_y\ra_{b,\,w}\equiv
{_b\la w|a^+_xa_xa^+_ya_y|w\ra_b}=
\sum_{v\leq v_F}|\la x|v\ra|^2
\sum_{v\leq v_F}|\la y|v\ra|^2 \non
-\left|\sum_{v \leq v_F} \la v|x\ra \la y|v\ra \right|^2 ,
\eea
where $v_F$ stands for the label of eigenvectors $|v\ra$,
such that there is exactly $N_e$ eigenvectors with $v \leq v_F$.
Consequently, all the correlation functions defined in this
section can be expressed in terms of the site-components $\la x|v\ra $
of the eigenvectors of $h_b(w)$.
For some low-period ion configurations the eigenproblem can be solved
exactly, while for any other ground-state
configurations of interest, and arbitrary boundary conditions,
one can resort to numerical exact-diagonalization procedures.

\section{Correlation functions -- exact results}

In this section, we calculate exactly ground-state correlation
functions of a finite chain with the periodic boundary conditions
only (so without any risk of confusion the subscript $p$ by the averages
can be dropped), and then the infinite-chain limits, in two cases.

First, for completeness, when $\rho_e \in [0,1]$ and $\rho_i =0$.
For such a particle densities,
$G$ consist of a single ion configuration,
called the {\em empty configuration}, for any $U$.
In this configuration, $g_x=0$ for all $x$, so we denote it $g \equiv 0$.

Second, when $\rho_e=1/2$ and $\rho_i =1/2$,
and $G=[1/2 ]_1$ consists of the two {\em checkerboard configurations}:
$\mbox{ch}^1=10$  and $\mbox{ch}^2=01$.
This holds also for any nonzero $U$, i.e. $\Delta U_{[1/2]_1}= (0,\infty)$
\cite{kennlieb}.

When $g_x=0$ for all $x$, all the non-vanishing matrix elements of $h_p(0)$
are given by
\bea
\label{hp0}
\la x|h_p(0)|x+1\ra =-t_{p,\,x}, \ \ \ \ t_{p,\,x}=1, \ \ \
\ \  x=0,1,\ldots,L-1 ,
\eea
Let $k \in \{2\pi l/L: \ l=0,1,\ldots,L-1 \}$, so that for any $L$,
$0 \leq k < 2\pi$.
The eigenvectors of $h_p(0)$, denoted $|k\ra$,
are specified by their site components, $\la x|k\ra$,
\bea
\label{xk}
\la x|k\ra=L^{-1/2}\mbox{e}^{ikx},
\eea
while the corresponding eigenvalues, $\varepsilon_k$, read
\bea
\label{ek}
\varepsilon_k=-2\cos k.
\eea
Introducing $k_F=\pi (N_e - 1)/L$, we obtain
\bea
\label{a+a0}
\la a^+_xa_{x+l} \ra_0=
\frac{1}{L}
\sum_{\stackrel{k\leq k_F,} {\scriptscriptstyle{k \geq 2\pi-k_F} }}
\exp{(i k l)} = \frac{1}{L} [ 1+2\sum_{0 < k \leq k_F} \cos(k l) ]
=\frac{1}{L} \frac{\sin(\pi l N_e/L)}{\sin(\pi l/L)}.
\eea
Thus, in the infinite-chain limit
\bea
\label{a+a0infty}
\lim_{L \to \infty} |\la a^+_xa_{x+l}\ra_0|^2 =
\frac{1}{\pi^2}\,
\frac{\sin^2(\pi \rho_el)}{l^2} ,
\eea
\bea
\label{nn0}
\lim_{L \to \infty} \la n_xn_{x+l}\ra_0 =
\rho_e^2-\frac{1}{\pi^2}\, \frac{\sin^2(\pi\rho_el)}{l^2}.
\eea
The analogous expressions for continuous counterparts of our system can
be found in \cite{martin, mahan}. The case of open
boundary conditions has been considered in detail in \cite{penson}.
Let us mention that the expression for the density-density correlation function $\la
n_xn_{x+l}\ra$, obtained in \cite{penson}, differs from our result (\ref{nn0}),
since the limit ${L\to \infty}$ is constructed in a different way,
namely as the limit of half-infinite chain (only one end becomes remote from
a fixed site).

\begin{figure}[t]
\centerline{\includegraphics[clip,width=8cm]{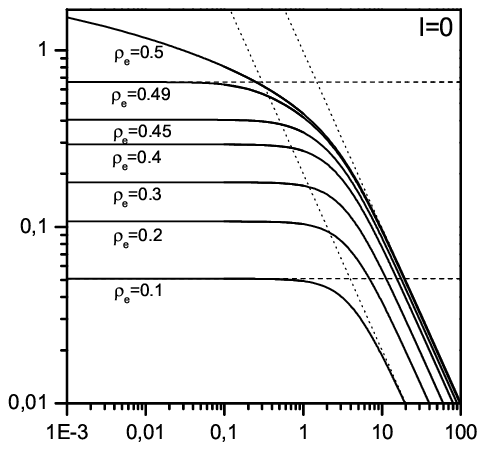}} 
\vspace{-13mm}
\centerline{\includegraphics[clip,width=8cm]{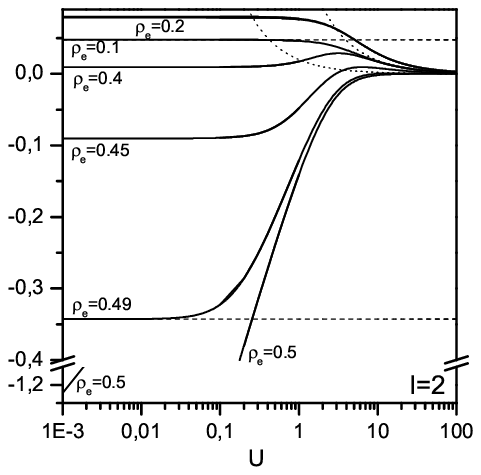}}
\caption[]
{Continuous lines:
$\lim_{L \to \infty} \tau_0 (l)$, given by (\ref{tau0infty}), versus $U$,
for $l=0,2$ and different values of $\rho_e$.
Dashed lines: the small $U$ asymptotes. Dotted lines: the large $U$ asymptotes.}
\label{fig1}
\end{figure}

More interesting is, of course, the case of the checkerboard
ground-state configurations $g=\mbox{ch}^1,\ \mbox{ch}^2$,
which are related by the primitive translation of the chain.
In this case, all the non-vanishing matrix elements of $h_p(\mbox{ch}^1)$
are given by
\bea
\label{hpch}
\la x|h_p(\mbox{ch}^1)|{x+1}\ra =-t_{p,\,x} = -1, \ \ \
\la x|h_p(\mbox{ch}^1)|x \ra = -U\mbox{ch}^1_x .
\eea
In the  basis $\lb |k_0\ra, |k_0+\pi\ra, |k_1\ra,
|k_1+\pi\ra,\ldots \rb$, $k=2\pi l/L$, $l=0,1,\ldots, L/2-1$,
obtained by reordering the basis $\lb |k\ra \rb$,
the matrix of $h_p(\mbox{ch}^1)$ becomes block-diagonal.
The diagonal blocks are 2 by 2 matrices
$\lbr h_p(\mbox{ch}^1) \rbr_k$ that in the basis
$\lb |k\ra , |k+\pi\ra \rb$  assume the form
\bea
\label{hpchk}
\lbr h_p(\mbox{ch}^1) \rbr_k = \lbr \begin{array}{cc}
\varepsilon_k-\frac{U}{2} & -\frac{U}{2} \non -\frac{U}{2} &
-\varepsilon_k-\frac{U}{2}\end{array} \rbr.
\eea
Thus, we obtain easily  the eigenvalues, $\Lambda_{\pm}(k)$,
\bea
\label{evach}
\Lambda^{\pm}(k)=-\frac{U}{2}\pm\Delta, \ \ \ \ \ \
\Delta=\sqrt{\varepsilon^2_k+\alpha^2}, \ \ \ \ \
\alpha=\frac{U}{2},
\eea
and the corresponding eigenvectors, $|k\ra^{\pm}_{ch}$, given by:
\bea
\label{evevch}
\lbr h_p(\mbox{ch}^1) \rbr_k
|k\ra^{\pm}_{ch}=\Lambda^{\pm}(k)|k\ra^{\pm}_{ch},
\eea
and
\bea
\label{evch}
|k\ra^{\pm}_{ch}=
(\gamma^{\pm}_k)^{-1} \ld \alpha|k\ra +\beta^{\pm}_k |k+\pi\ra \rd , \ \ \ \
\beta^{\pm}_k=\varepsilon_k\mp \Delta,
\eea
where the normalizing factor $\gamma^{\pm}_k$ is
\bea
\label{k1norm}
(\gamma^{\pm}_k)^2\equiv
\alpha^2+(\beta^{\pm}_k)^2=2\Delta\beta^{\pm}_k.
\eea
Since in what follows, we shall be interested in $\rho_e \leq 1/2$ only,
we can restrict our considerations to the lowest $L/2$
eigenvalues $\Lambda^-(k)$ and the corresponding eigenvectors
$|k\ra^-$, and set
\bea
\label{notation}
|k\ra_{ch}\equiv|k\ra^-_{ch}, \ \ \
\beta_k\equiv\beta^-_k=\varepsilon_k+\Delta, \ \ \
\gamma_k\equiv\gamma^-_k .
\eea
Then, the site-components of eigenvectors $|k\ra_{ch}$ read
\bea
\label{xk1}
\la x |k\ra_{ch}=\frac{1}{\sqrt{L}}\,\frac{1}{\gamma_k}\ld
\alpha \mbox{e}^{ikx}+\beta_k \mbox{e}^{i(k+\pi)x} \rd,
\eea
and consequently, setting $\la \mbox{ch}^1|a^+_xa_x|\mbox{ch}^1\ra
\equiv \la a^+_xa_x\ra_{1}$, the local density $\la n_x \ra_{1}$ reads
\bea
\label{nx1}
\la n_x \ra_{1} = \la a^+_xa_x\ra_{1}
=\frac{1}{L}\sum_{\stackrel{k\leq k_F,} {\scriptscriptstyle{k \geq \pi-k_F} }}
|\la x|k\ra_{ch}|^2
=\rho_e+(-1)^x \alpha\tau_0(0),
\eea
where
\bea
\label{tau0l}
\tau_0(l) \equiv \frac{1}{L}\lb\frac{1}{\sqrt{4+\alpha^2}}
+ 2 \sum_{0 < k \leq k_F} \frac {\cos(k l)}{\Delta} \rb.
\eea
In the thermodynamic limit,
\bea
\label{tau0infty}
\lim_{L \to \infty} \tau_0(l)=\frac{1}{\pi}
\int\limits_0^{ \pi \rho_e} \mbox{d}k \frac{\cos(k l)}{\Delta}\non
 =\frac{1}{\pi}\int\limits_0^{\pi\rho_e}\mbox{d}k
\frac{\cos(k l)}{\sqrt{\alpha^2+\varepsilon_k^2}}
=\frac{\kappa}{2\pi}\int\limits_0^{\pi\rho_e}\mbox{d}k
\frac{\cos(k l)}{\sqrt{1-\kappa^2\sin^2k}}\,
,  \ \ \ \ \ \rho_e\leq\frac{1}{2},\ \ \
\kappa^2=\frac{1}{1+(\frac{U}{4})^2} < 1 .
\eea
Since, the function $\lim_{L \to \infty} \tau_0(l)$ appears in
many expressions for correlation functions, in the sequel,
it is interesting to see its dependence on $\rho_e$ and $U$.
This is shown in Fig.~\ref{fig1}.
The analytic expressions for the asymptotic behaviour of this function,
for $U \to 0$ and for $U \to \infty$, are given in Appendix B.
Using these asymptotic formulae we find the corresponding asymptotic
expressions for the local density $\la n_x \ra_{1}$:
for $U \to 0$
\bea
\lim_{L\to \infty} \la n_x \ra_1 \approx
\lb \begin{array}{ll} \rho_e+(-1)^x{\alpha}{(2\pi)^{-1}}
\ln \tan\lbr\pi(1+2\rho_e)/4 \rbr,  & \rho_e < \frac{1}{2}, \\
\rho_e+(-1)^x{\alpha}{(2\pi)^{-1}}\ld a_0+2\ln2-\ln\alpha \rd, & \rho_e=\frac{1}{2},
\end{array} \right.
\eea
while for $U \to \infty$ and any $\rho_e \leq 1/2$,
\bea
\lim_{U\to\infty}\lim_{L\to\infty} \la n_x \ra_1 =\rho_e+(-1)^x\rho_e ,
\eea
which in the case of half-filling (i.e. for $\rho_e =1/2$) amounts to ch$^1$.
Moreover, in the case of half-filling,
the local electron-density of an infinite chain can be expressed by
the complete elliptic integral of the first kind $K$ \cite{as},
at the point $\kappa^2$,
\bea
\label{nx1hf}
\la n_x \ra_{1} =\frac{1}{2}+(-1)^x \frac{U\kappa}{4\pi}\,
\int\limits_0^{\pi/2}
\frac{\mbox{d}k}{\sqrt{1-\kappa^2\sin^2k}}=
\frac{1}{2}+(-1)^x\frac{U\kappa}{4\pi}K(\kappa^2),\ \ \rho_e=\frac{1}{2}.
\eea

Now, consider the two-point correlation function
$ \la a^+_xa_{x+l} \ra_1 \equiv
\la \mbox{ch}^1 | a^+_xa_{x+l}|\mbox{ch}^1 \ra_{ch}$.
For $\rho_e \leq 1/2$, we find
\bea
\label{a+a1}
\la a^+_xa_{x+l}\ra_{1} =
\sum_{\stackrel{k\leq k_F,} {\scriptscriptstyle{k \geq \pi-k_F} }}
{_{1}\la k|x\ra \la x+l|k \ra_1}
 \non =
\frac{1}{L} \sum_{\stackrel{k\leq k_F,} {\scriptscriptstyle{k \geq \pi-k_F}} }
\frac{\mbox{e}^{ikl}}{\gamma_k^2}\lb
\alpha^2+(-1)^l\beta_k^2+(-1)^x\alpha\beta_k\lbr 1+(-1)^l \rbr \rb ,
\eea
which for nonzero and even $l$ assumes the form
\bea
\label{a+a1even}
 l=2s, \ \ s=1,2,\ldots , \ \ \ \ \
 \la a^+_xa_{x+l}\ra_{1}
 =\frac{1}{L} \sum_{\stackrel{k\leq k_F,} {\scriptscriptstyle{k \geq \pi-k_F}}}
\mbox{e}^{ikl}\ld  1+(-1)^x \frac{\alpha}{\Delta} \rd \non
=\la a^+_xa_{x+l} \ra_0 +
(-1)^x \alpha
\frac{1}{L}\lb\frac{1}{\sqrt{4+\alpha^2}}
+ 2\!\! \sum_{0 < k \leq k_F}\!\! \frac {\cos(k l)}{\Delta} \rb
\equiv \la a^+_xa_{x+l} \ra_0
+ (-1)^x\alpha\tau_0(2s),
\eea
while for odd $l$, the form:
\bea
\label{a+a1odd}
\la a^+_xa_{x+l}\ra_{1}
=\frac{1}{L} \sum_{\stackrel{k\leq k_F,} {\scriptscriptstyle{k \geq \pi-k_F}}}
\mbox{e}^{ikl}\ld  - \frac{\varepsilon_k}{\Delta} \rd
=
\tau_1(l),
\ \ \ \ l=2s+1, \ \  s=0,1,\ldots ,
\eea
where
\bea
\label{tau1l}
\tau_1(l) \equiv \frac{1}{L} \lb \frac{2}{\sqrt{4+\alpha^2}} +2 \sum_{0 < k \leq k_F}
\cos(k l) \ld  - \frac{\ve_k}{\Delta} \rd \rb .
\eea
We note that the functions $\tau_0$ and $\tau_1$ satisfy the relation:
\bea
\label{tau0tau1}
\tau_1(2s+1)=\tau_0(2s+2)+\tau_0(2s).
\eea
Summarizing, for $\rho_e \leq 1/2$,
the two-point function $\la a^+_xa_{x+l} \ra_1$ reads
\bea
\label{a+a1final}
\la a^+_xa_{x+l} \ra_1 =
\lb \begin{array}{lll} \tau_1(l), & l=2s+1, & s=0,1,\ldots ,\\
\la a^+_xa_{x+l} \ra_0 + (-1)^x\alpha\tau_0(l), & l=2s, & s=1,2,\ldots .
\end{array}  \right.
\eea
As expected, since for $U \to 0$: $\tau_1(l) \to \la a^+_xa_{x+l}\ra_0$,
and $\alpha \lim_{L\to\infty} \tau_0(l)\to 0$ (see Appendix B),
the two-point function
$\la a^+_xa_{x+l}\ra_1 \to \la a^+_xa_{x+l}\ra_0$, as $U \to 0$.
On the other hand, using the large $U$ asymptotic formulae for
$\lim_{L\to\infty} \tau_0(l)$ (see Appendix B),
we obtain: for $U \to \infty$
\bea
\lim_{L\to\infty} \la a^+_xa_{x+l} \ra_1 \approx
\lb \begin{array}{lll} ({\pi\alpha(l+1)})^{-1} {\sin\pi \rho_e(l+1)} & & \\
+({\pi\alpha(l-1)})^{-1}{\sin\pi \rho_e(l-1)}, & l=2s+1, & s=0,1,\ldots, \\
\ld 1+(-1)^x \rd (\pi l)^{-1}{\sin\pi\rho_el}, & l=2s, & s=1,2,\ldots .
\end{array} \right.
\eea
In particular, for $\rho_e=1/2$,
$\lim_{U \to \infty} \lim_{L \to \infty} \la a^+_xa_{x+l}\ra_1 \equiv  0$.
From now on, we consider only the case of $\rho_e=1/2$.
In the infinite-chain limit,
$\lim_{L \to \infty}\la a^+_xa_{x+2s} \ra_0 =0$, $s=1,2,\ldots$, thus
\bea
\label{a+a1finalinfty}
\lim_{L \to \infty}\la a^+_xa_{x+l} \ra_1 =
\lb \begin{array}{lll} \lim_{L \to \infty} \tau_1(l), & l=2s+1, & s=0,1,\ldots , \\
(-1)^x\alpha \lim_{L \to \infty}\tau_0(l), & l=2s, & s=1,2,\ldots .
\end{array}  \right .
\eea
By \cite{pbm} (see also Appendix B),
\bea
\label{tau0eveninf}
\lim_{L \rightarrow \infty} \tau_0(2s)=
\frac{\kappa}{4\pi}\int\limits_0^{\pi}
\frac{\cos(2sk)}{\sqrt{1-\kappa^2\sin^2k}}\mbox{d}k
=(-1)^s\frac{\kappa}{4\pi}\int\limits_0^{\pi}
\frac{\cos(2sk)}{\sqrt{1-\kappa^2\cos^2k}}\mbox{d}k
\nonumber \\
=(-1)^s\frac{\kappa^{2s+1}}{2^{3s+2}}\,\frac{(2s-1)!!}{s!}\, F\ld
s+\frac{1}{2},s+\frac{1}{2},2s+1;\kappa^2 \rd ,
\eea
and
\bea
\label{tau1oddinf}
\lim_{L \rightarrow \infty} \tau_1(2s+1)=
(-1)^s\frac{\kappa^{2s+1}}{2^{3s+2}}\,\frac{(2s-1)!!}{s!}\, \lb F\ld
s+\frac{1}{2},s+\frac{1}{2},2s+1;\kappa^2 \rd  \right. \non
\left. -\frac{\kappa^2}{2^3}\,\frac{2s+1}{s+1} F\ld
s+\frac{3}{2},s+\frac{3}{2},2s+3;\kappa^2 \rd \rb ,
\eea
where $F(a,b,c;z)$  stands for the hypergeometric function \cite{as, be}.
By means of an integral representation of $F(a,b,c;z)$ \cite{be}, $\tau_0(2s)$
can be rewritten in the following form
\bea
\label{tau0evenLap}
\lim_{L \rightarrow \infty} \tau_0(2s)
=(-1)^s\,\frac{\kappa^{2s+1}}{4\pi} \int\limits_0^1 \exp \lbr {s\ln
\frac{t(1-t)}{1-\kappa^2t} } \rbr
\frac{\mbox{d}t}{\sqrt{t(1-t)(1-\kappa^2t)}},
\eea
which is suitable for determining the asymptotics of $\tau_0(2s)$
for $s \rightarrow \infty $. Using the Laplace asymptotic formula
\cite{bo} for the integral in (\ref{tau0evenLap}), we obtain
\bea
\label{tau0evenasymp}
\lim_{L \rightarrow \infty} \tau_0(2s)\approx
(-1)^s \frac{1}{4\sqrt{\pi}}\,
\frac{1}{\sqrt[4]{(\frac{U}{4})^2(1+(\frac{U}{4})^2)}}\,
\frac{\exp({-{s}/{\xi}})}{\sqrt{s}} ,
\eea
where the {\em correlation length} $\xi$ is defined by
\bea
\label{ksi}
\xi\equiv-\frac{1}{2\ln \ld \sqrt{1+(\frac{U}{4})^2}- \sqrt{(\frac{U}{4})^2} \rd} ,
\eea
and has the following asymptotic behaviour:
\bea
\label{ksiasympsmall}
\xi \approx \frac{2}{U}, \ \ \ \mbox{for}\ \  U \rightarrow 0,
\eea
and
\bea
\label{ksiasymplarge}
\xi \approx \frac{1}{2\ln(\frac{U}{2})}, \ \ \ \mbox{for} \ \
U \rightarrow \infty.
\eea

\begin{figure}[t]
\centerline{\includegraphics[clip,width=8cm]{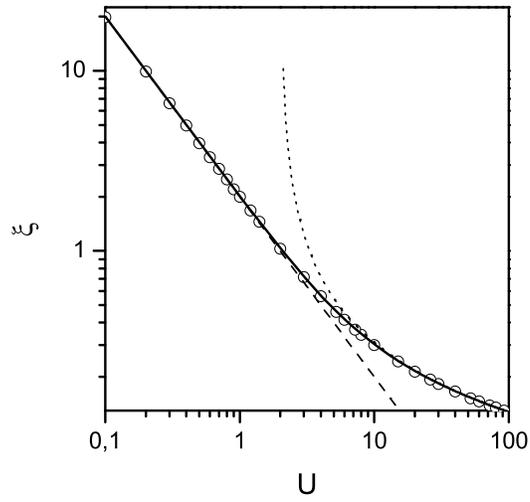}}
\caption[]
{Continuous line: the exact correlation length for the checkerboard configurations,
given by (\ref{ksi}), versus $U$.
Dashed line: the small $U$ asymptote (\ref{ksiasympsmall}).
Dotted line: the large $U$ asymptote (\ref{ksiasymplarge}).
Open circles: $\xi$ obtained from fitting numerically calculated
correlation function $\la a^+_xa_{x+2s} \ra_{1}$
with the formula $const \exp(-s/\xi)$.}
\label{fig2}
\end{figure}

\begin{figure}[t]
\centerline{\includegraphics[clip,width=8cm]{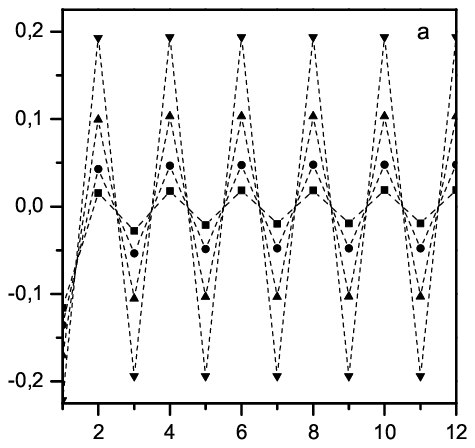}}
\vspace{-10mm}
\centerline{\includegraphics[clip,width=8cm]{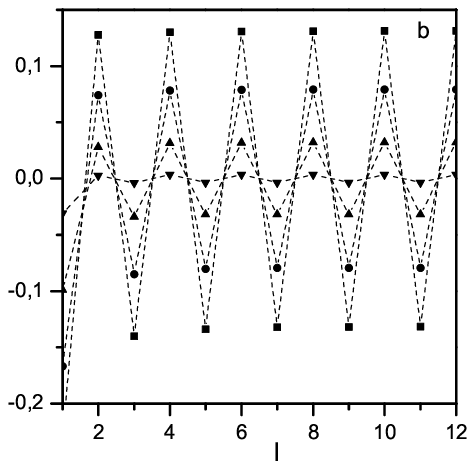}}
\caption[]
{(a) ${\cal{L}}(l)$, given by (\ref{Lplchessinfty}), versus $l$, for $G=[1/2]_1$,
and different values of $U$.
(b) ${\cal{S}}(l)$, given by (\ref{Splchessinfty}), versus $l$, for $G=[1/2]_1$,
and different values of $U$.
Filled triangles with base up: $U=5$, filled triangles with base down: $U=2$,
filled circles: $U=1$, filled squares: $U=0.5$.}
\label{fig3}
\end{figure}

\begin{figure}[t]
\centerline{\includegraphics[clip,width=8cm]{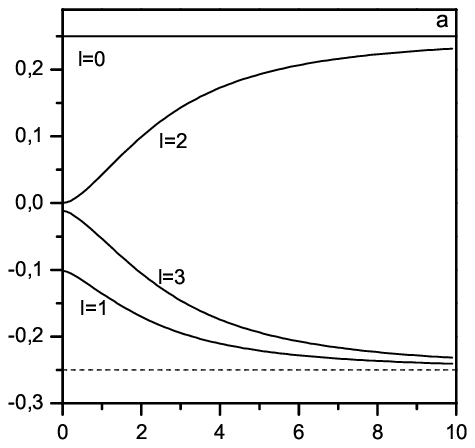}}
\vspace{-10mm}
\centerline{\includegraphics[clip,width=8cm]{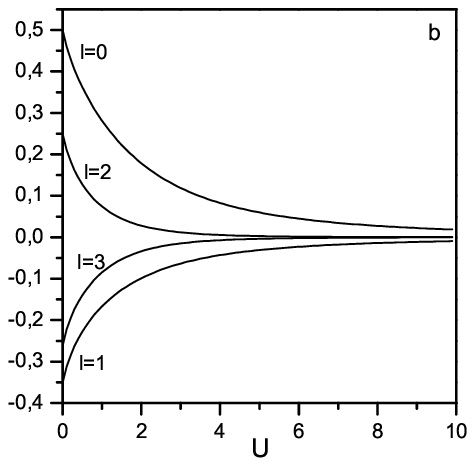}}
\caption[]
{(a) ${\cal{L}}(l)$, given by (\ref{Lplchessinfty}), versus $U$, for $G=[1/2]_1$,
and different values of $l$.
(b) ${\cal{S}}(l)$, given by (\ref{Splchessinfty}), versus $U$, for $G=[1/2]_1$,
and different values of $l$.
In the scale of the figure, the plots for odd $l\geq 5$ cannot be
distinguished from the plot for $l=3$.
Similarly, the plots for $l\geq 4$ cannot be
distinguished from the plot for $l=2$.}
\label{fig4}
\end{figure}

\begin{figure*}[t]
\centerline{\includegraphics[clip,width=8cm]{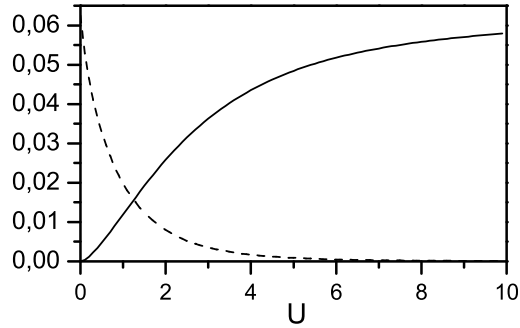}}
\caption[]
{Continuous line: ${\cal{L}}$ for $G=[1/2]_1$, given by (\ref{Lchess}), versus $U$.
Dashed line: ${\cal{S}}$ for $G=[1/2]_1$, given by (\ref{Schess}), versus $U$.}
\label{fig5}
\end{figure*}

In Fig.~\ref{fig2} we show $\xi$ as a function of $U$,
obtained from the exact formula (\ref{ksi}) and from fitting  numerically
calculated correlation function $\la a^+_xa_{x+2s} \ra_{1}$
with the formula $const \exp(-s/\xi)$.
We plot also in Fig.~\ref{fig2} the asymptotic formulae (\ref{ksiasympsmall}) and
(\ref{ksiasymplarge}).
As a matter of fact, the formula (\ref{ksiasympsmall}) approximates
the exact value of $\xi$ with accuracy $10^{-2}$, up to $U=1$.
The formula (\ref{ksiasymplarge}) approximates $\xi$ with accuracy
$2,5\cdot 10^{-2}$, for $U \geq 10$.
Consequently, we find the asymptotic forms of the two-point function
$\la a^+_xa_{x+l} \ra_{1}$:
\bea
\label{a+a1evenasymp}
l=2s,\ \ \ \
|\la a^+_xa_{x+2s} \ra_{1} |^2 \approx \frac{1}{2\pi}\,
\frac{\frac{U}{4}}{\sqrt{1+(\frac{U}{4})^2}}\,
\frac{\exp(-{2s}/{\xi})}{2s},
\eea
and
\bea
\label{a+a1oddasymp}
l=2s+1,\ \
|\la a^+_xa_{x+2s+1} \ra_{1} |^2 \approx \frac{1}{8\pi}\,
\frac{1}{\sqrt{(\frac{U}{4})^2(1+(\frac{U}{4})^2)}}\, \lbr 1-
\frac{\exp (-1/\xi)}{\sqrt{1+{1}/{s}}} \rbr^2 \,
\frac{\exp(-{2s}/{\xi})}{2s} .
\eea
Having obtained the correlation functions $\la n_x \ra_{1}$ and
$\la a^+_xa_{x+l} \ra_{1}$, we can easily calculate other correlations
of interest. The short-range correlation functions (\ref{Lbx}), (\ref{Sbx}), for $l=0$ read:
\bea
\label{Lp0}
{\cal{L}}_p(0)=\rho_e(1-\rho_e),
\eea
\bea
\label{Sp0}
{\cal{S}}_p(0)=\frac{1}{2}-\alpha\tau_0(0).
\eea
Then, for $l > 0$ and any $\rho_e \leq 1/2$ we find
\bea
\label{Lplchess}
{\cal{L}}_p(l)= (-1)^l\alpha^2\tau^2_0(0)-
\lb \begin{array}{lll} \tau_1^2(l), & l=2s+1, & s=0,1,\ldots , \\
\la a^+_xa_{x+l}\ra_0^2+\alpha^2\tau_0^2(l), & l=2s, &  s=1,2,\ldots ,  \end{array} \right.
\eea

\bea
\label{Splchess}
{\cal{S}}_p(l)=(\rho_e-\frac{1}{2})^2 \non
+ (-1)^l\lbr \frac{1}{2} -\alpha\tau_0(0) \rbr^2
-\lb \begin{array}{lll}
\tau_1^2(l), &  l=2s+1, & s=0,1,\ldots , \\
\la a^+_xa_{x+l}\ra_0^2+\alpha^2\tau_0^2(l), & l=2s, & s=1,2,\ldots . \end{array}
\right.
\eea
In the infinite-chain limit and for $\rho_e=1/2$, the above short-range correlation
functions assume the form:
\bea
\label{Lplchessinfty}
{\cal{L}}(l)=\lim_{L \to \infty} {\cal{L}}_p(l) \non
=(-1)^l\alpha^2\lim_{L\to\infty}\tau^2_0(0)-
\lb \begin{array}{lll}
\lim_{L\to\infty} \tau^2_1(l), & l=2s+1, & s=0,1\ldots , \\
\alpha^2\lim_{L\to\infty} \tau^2_0(l), & l=2s, & s=1,2\ldots ,
\end{array} \right.
\eea

\bea
\label{Splchessinfty}
{\cal{S}}(l)= \lim_{L\to\infty}{\cal{S}}_p(l) \non
=(-1)^l\lbr \frac{1}{2} -\alpha\lim_{L\to\infty}\tau_0(0) \rbr^2
-\lb \begin{array}{lll}
\lim_{L\to\infty}\tau_1^2(l), &  l=2s+1, & s=0,1,\ldots , \\
\alpha^2\lim_{L\to\infty}\tau_0^2(l), & l=2s, & s=1,2,\ldots . \end{array}
\right.
\eea
The short-range correlations, given by (\ref{Lplchessinfty})
and (\ref{Splchessinfty}), are shown
in Fig.~\ref{fig3} as functions of the distance, and in Fig.~\ref{fig4} as functions of $U$.
As $U$ increases, the absolute value of ${\cal{L}}$ increases,
while the absolute value of ${\cal{S}}$ decreases. At distance $l=1$,
a minimum, known as the correlation hole \cite{mahan}, is clearly visible.
Then, at distances greater than a few lattice constants,
the short-range correlations become periodic, due to the exponential decay
of the two-point function $\la a^+_xa_{x+l} \ra_{1}$.
Using (\ref{P}), (\ref{L}), and (\ref{S}), we easily find the
infinite-chain long-range correlation functions ${\cal{P}}(\pi)$, ${\cal{L}}$,
and ${\cal{S}}$. For $\rho_e=1/2$ they read:
\bea
\label{Pchess}
{\cal{P}}(\pi)= \lim_{L\to \infty} {\cal{P}}_p (\pi) =
\alpha^2 \lim_{L\to \infty} \tau^2_0(0) ,
\eea
\bea
\label{Lchess}
{\cal{L}}= \lim_{L \to \infty} {\cal{L}}_p =
\frac{1}{4} \alpha^2 \lim_{L\to \infty} \tau^2_0(0) ,
\eea
\bea
\label{Schess}
{\cal{S}}=\frac{1}{16} \ld 1-2\alpha \lim_{L\to\infty}\tau_0(0) \rd^2 .
\eea
For checkerboard configurations, ${\cal{P}}(\pi)$ and ${\cal{L}}$
are proportional.
Using the asymptotic formulae of Appendix B, we find that as $U \to \infty$,
${\cal{P}}(\pi) \to 1/4$, ${\cal{L}} \to 1/16$, and ${\cal{S}} \to 0$.
The variation of long-range correlations versus $U$ is shown in Fig.~\ref{fig5}.
As $U$ increases, the quantum fluctuations decrease, but rather slowly,
disappearing only in the limit of infinite $U$.
For $U=5$, ${\cal{L}}$  attains 77\% of its saturation value,
and for $U=8.3$, 90\% of it.

\section{Correlation functions -- numerical results}

In the previous section we have been able to determine analytically
the spatial and $U$ dependence of a number of correlation functions,
in the case of $G=[1/2]$.
Here, our aim is to present the spatial and $U$ dependence of the same
kind of correlation functions, but with other choices of the set $G$.
Specifically, we consider the following sets of low-period ground-state
configurations: $[1/6]_1$, $[1/6]_2$, $[1/6]_3$, $[1/3]$, $[2/5]$, $[3/7]$,
$[1/4]_2$; atomic, 2-molecular and 3-molecular configurations are included
in the list. The corresponding stability intervals of $U$,
determined approximately by means of restricted ground-state phase diagrams
obtained in \cite{gjl}, read: $\Delta U_{[1/6]_1}= [0.6 , \infty )$,
$\Delta U_{[1/6]_2}= [0.3 , 2.7]$, $\Delta U_{[1/6]_3}= [0.1 , 1.0 ]$,
$\Delta U_{[1/3]_1}= [0.2 , \infty )$, $\Delta U_{[2/5]_1}= [0.1 , \infty )$,
$\Delta U_{[3/7]_1}= [0.05 , \infty )$, $\Delta U_{[1/4]_2}= [0.3 , 2.2]$.
In contradistinction to the previous section, the results reported
here have been obtained numerically. We have used numerical procedures elaborated
for a study of spin chains \cite{derkro}, with the main part being an exact
diagonalization of the matrix $\la x|h_b(w)|y \ra$, defined by (\ref{h}), (\ref{hxy}).
The size of the chains has varied from a few hundred up to a few thousand sites.
In the calculations, we have imposed the free boundary conditions on
a chain whose number of sites, $L$, was chosen in such a way that $L$ was
a multiple of $4q$, with $q$ being the period of the configurations
in $G$ under consideration. The one-point, broken symmetry averages
$\la n_x \ra_{b,\,g}$,
and the two-point, symmetric, short-range correlations,
${\cal{L}}_{b,\,x}(l)$ and ${\cal{S}}_{b,\,x}(l)$,
have been calculated using the formulae (\ref{nxny}).
To weaken the effect of the chain ends, the site $x$ and distance $l$
have been chosen so that all the sites involved in the calculations of
correlations were well inside the chain.
The results obtained for chains of different sizes have been extrapolated
to the infinite-chain limit. Practically, it has appeared that the results
obtained for chains of $L=420$ coincided with the corresponding infinite-chain
results with accuracy (the relative error) exceeding $10^{-12}$.

\begin{figure*}[t]
\centerline{\includegraphics[clip,width=8cm]{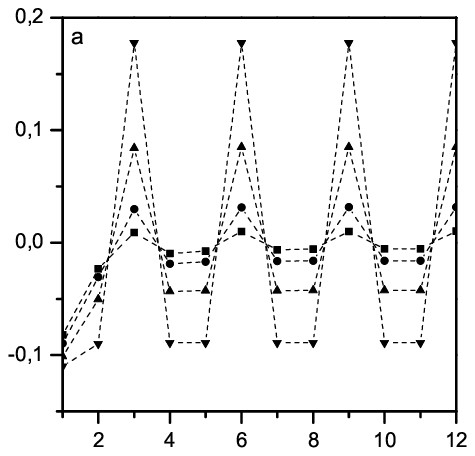}
\hfill
\includegraphics[clip,width=8cm]{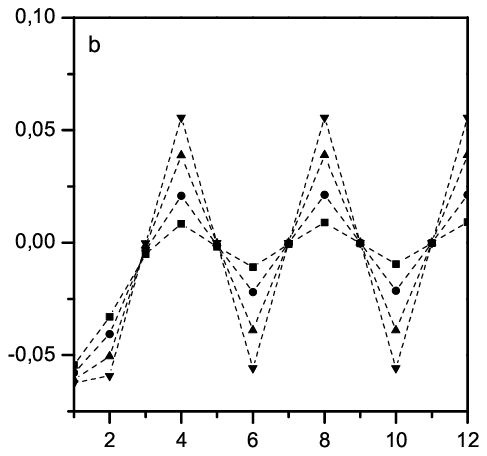}}
\vspace{-15mm}
\centerline{\includegraphics[clip,width=8cm]{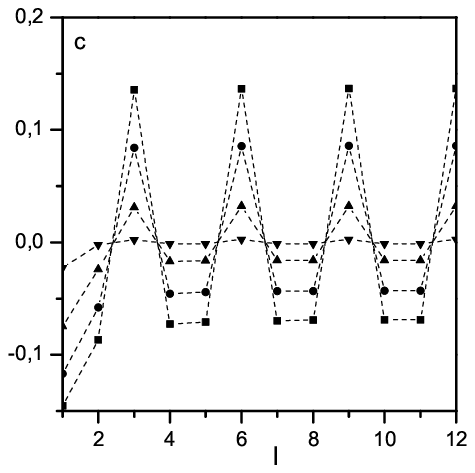}
\hfill
\includegraphics[clip,width=8cm]{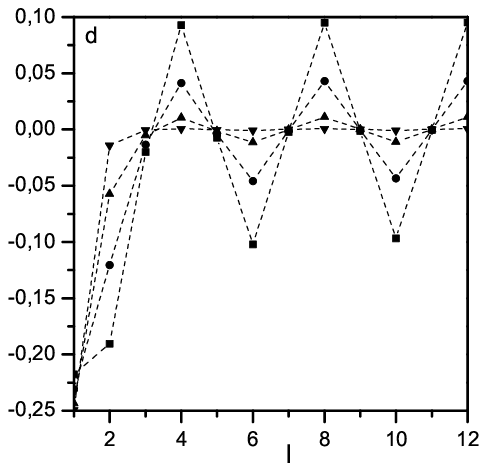}}
\caption[]
{Infinite chain limit of ${\cal{L}}_{f,x}(l)$, given by (\ref{Lbx}),
versus $l$, for $G=[1/3]_1$ (a), $G=[1/4]_2$ (b), and for different values of $U$.
Infinite chain limit of ${\cal{S}}(l)$, given by (\ref{Sbx}),
versus $l$, for $G=[1/3]_1$ (c), $G=[1/4]_2$ (d), and different values of $U$.
Filled triangles with base up: $U=5$, filled triangles with base down: $U=2$,
filled circles: $U=1$, filled squares: $U=0.5$.}
\label{fig6}
\end{figure*}

\begin{figure*}[t]
\centerline{\includegraphics[clip,width=8cm]{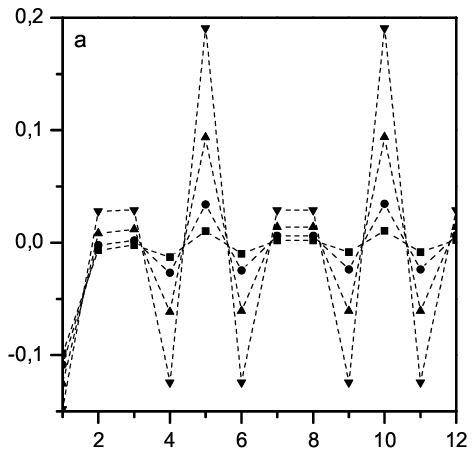}
\hfill
\includegraphics[clip,width=8cm]{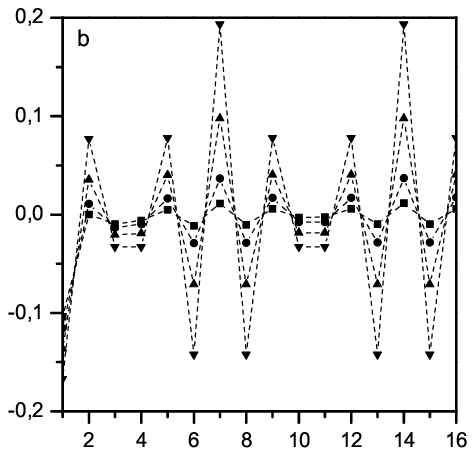}}
\vspace{-15mm}
\centerline{\includegraphics[clip,width=8cm]{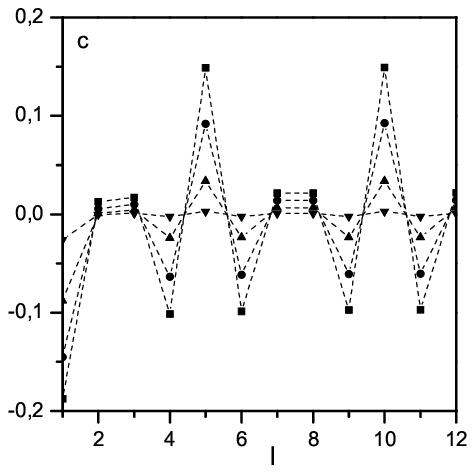}
\hfill
\includegraphics[clip,width=8cm]{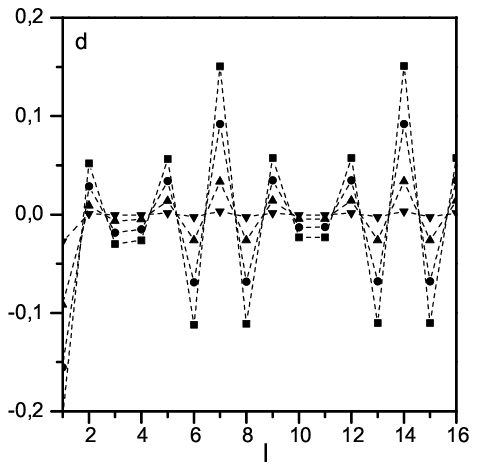}}
\caption[]
{Infinite chain limit of ${\cal{L}}_{f,x}(l)$, given by (\ref{Lbx}),
versus $l$, for $G=[2/5]_1$ (a), $G=[3/7]_1$ (b), and for different values of $U$.
Infinite chain limit of ${\cal{S}}(l)$, given by (\ref{Sbx}),
versus $l$, for $G=[2/5]_1$ (c), $G=[3/7]_1$ (d), and different values of $U$.
Filled triangles with base up: $U=5$, filled triangles with base down: $U=2$,
filled circles: $U=1$, filled squares: $U=0.5$.}
\label{fig7}
\end{figure*}

\begin{figure*}[t]
\centerline{\includegraphics[clip,width=8cm]{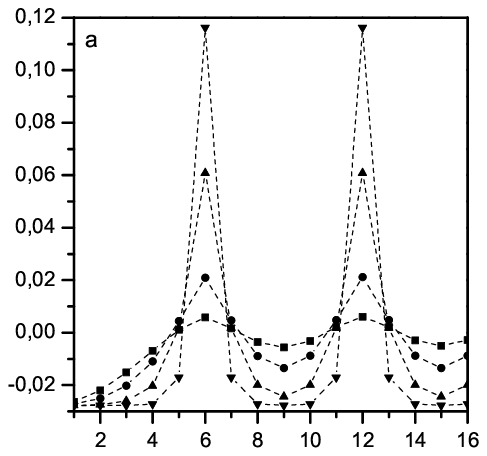}
\hfill
\includegraphics[clip,width=8cm]{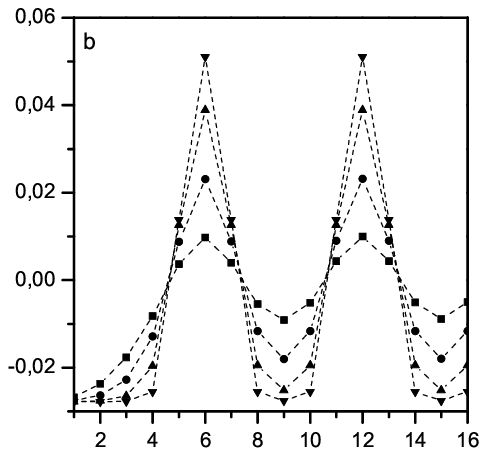}}
\vspace{-15mm}
\centerline{\includegraphics[clip,width=8cm]{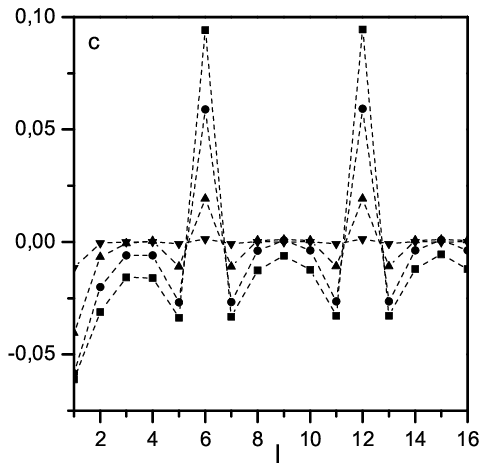}
\hfill
\includegraphics[clip,width=8cm]{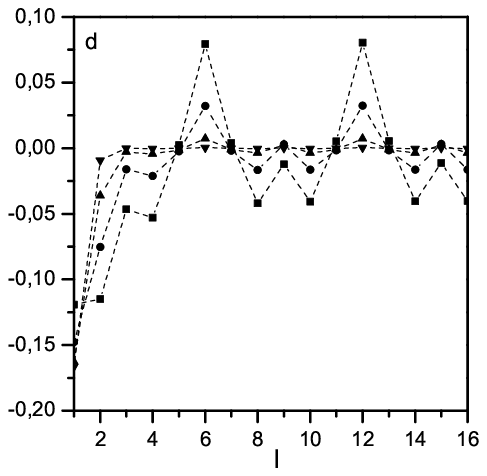}}
\caption[]
{Infinite chain limit of ${\cal{L}}_{f,x}(l)$, given by (\ref{Lbx}),
versus $l$, for $G=[1/6]_1$ (a), $G=[1/6]_2$ (b), and for different values of $U$.
Infinite chain limit of ${\cal{S}}(l)$, given by (\ref{Sbx}),
versus $l$, for $G=[1/6]_1$ (c), $G=[1/6]_2$ (d), and different values of $U$.
Filled triangles with base up: $U=5$, filled triangles with base down: $U=2$,
filled circles: $U=1$, filled squares: $U=0.5$.}
\label{fig8}
\end{figure*}

In Figs.~\ref{fig6}, \ref{fig7}, \ref{fig8}, we show short-range correlations
as functions of distance.
All the plots of two-point correlation functions, versus distance,
exhibit a minimum at distance one,
i.e. the so called correlation hole \cite{mahan}, and after
a few further steps become perfectly periodic, because of the exponential decay
of the function $\la a^+_xa_{x+l} \ra $ (see below).
In Figs.~\ref{fig9} -- \ref{fig12} we have plotted the
short-range correlations, for a few small distances, as functions of $U$.
Typically, in the presented cases the short-range correlations vary
monotonically with $U$. However, ${\cal{S}}(1)$ for $G=[1/6]_1$
(shown in Fig.~\ref{fig12})
exhibits a minimum for some $U < 1$. It is caused by a non-monotonic
behaviour of $\la n_x \ra_g$ at nearest neighbours of the occupied site;
for small $U$ ($U< 0.5$ roughly) it increases, and then decreases (see
Fig.~\ref{fig13}).
We note also that, as $U \to \infty$,
${\cal{S}}(0)$, ${\cal{S}}(1)$ do not tend to $0$, for molecular
configurations. This is the case also for ${\cal{S}}(2)$ in the case of
3-molecular configurations $[1/6]_3$. The reason is that for
molecular configurations, the local density of electrons at site $x$
does not tend to $g_x$ as $U \to \infty$, but obeys, for molecules greater than
$3$-molecules, some nontrivial distribution. This is shown in Fig.~\ref{fig13}.

Of particular interest is the spatial decay of the two-point correlation
function $|\la a^+_x a_{x+l} \ra_b|^2$.
The exact calculations carried out for checkerboard configurations,
in the previous section, suggest that for periodic ion configurations
this decay is of the form $l^{-\gamma} \exp(-l/\xi)$,
with some positive $\gamma$. Thus, this decay can be
characterized by the correlation length $\xi$ and the index $\gamma$.
Specifically, for checkerboard configurations we have obtained $\xi$
as a function of $U$ (\ref{ksi}), with
$\xi^{-1} = U/2$, for small $U$, that is $\xi^{-1}$ amounts to a half
of the gap at the Fermi level,
and with $\gamma =1$, according to  (\ref{a+a1evenasymp}),
(\ref{a+a1oddasymp}).
For $G$ other than the checkerboard configurations, we have approximated
the large distance behaviour of $|\la a^+_x a_{x+l} \ra_b|^2$ by the formula
$const \exp(-l/\xi)$, thus ignoring the possible power-law prefactor
$l^{-\gamma}$. The data are shown in Fig.~\ref{fig14} . Comparing the data obtained
in such calculations of $\xi$ with the exact results in the case of
the checkerboard configurations, we estimate the error to be at the level
of a few per cent. To reach such an accuracy, in the range of $U < 0.4$
it was necessary to consider a chain with $L=2\cdot 10^3$, while for larger
$U$ it was enough to take $L= 10^3$.
For all the studied periodic configurations, we have found that for small
$U$ the inverse correlation length $\xi^{-1} \approx const \Delta$,
with $\Delta$ being the gap at the Fermi level, and $\Delta \sim U$ .

Finally we present our results concerned with the defined in Section 3
long-range correlation functions. Our purpose was not only
to calculate the correlations ${\cal{P}}_f(k)$, ${\cal{L}}_f$,
and ${\cal{S}}_f$, in the infinite-chain limit, but also to
observe their dependence on the chain size $L$, in a finite chain with
the free boundary conditions imposed ($b=f$).
To calculate the long-range correlations, we used their definitions
(\ref{Pb}), (\ref{Lb}), (\ref{Sb}), which involve all sites of the chain,
and the simplified expressions, involving $L/2$ central sites of the chain,
\bea
\label{P'b}
{\cal{P}}'_f(k) = \frac{2}{L} \sum_{l=0}^{L/2 -1}
\mbox{e}^{-ikl} {\cal{L}}_{f,x}(l),
\eea
\bea
\label{L'b}
{\cal{L}}'_f = \frac{2}{L} \sum_{l=0}^{L/2 -1}
E_p(l) {\cal{L}}_{f,x}(l),
\eea
\bea
\label{S'b}
{\cal{S}}'_f = \frac{2}{L} \sum_{l=0}^{L/2 -1}
E_p(l) {\cal{S}}_{f,x}(l) ,
\eea
where $x=L/4$.
As shown in Fig.~\ref{fig15}, for particular sets $G$ and values of $U$,
the long-range correlations vary linearly with $1/L$, and the
simplified expressions, being less sensitive to the boundaries
of the chain, for given $L$ constitute a lot better approximation
to the infinite-chain value of correlations than the defining formulae.
The dependence of infinite-chain values of ${\cal{P}}$, ${\cal{L}}$,
and ${\cal{S}}$ on $U$, for different choices of the set $G$,
is displayed in Figs.~\ref{fig16},~\ref{fig17}. As a rule,
${\cal{L}}$ increases towards a positive saturation value, while ${\cal{S}}$
decreases towards zero, for all the considered sets $G$,
except $G=[1/6]_3$, when ${\cal{S}}$ exhibits a minimum at $U \approx 2$.
For $U > 2$, ${\cal{S}}$ increases towards a positive asymptote.
An inspection of the local density of electrons, as a function $U$, reveals a
non-monotonic behaviour of this density at nearest neighbours of the central
occupied site of the $3$--molecule: it has the maximum at $U \approx 2$ and decreases
towards a positive asymptote, as $U \to \infty$ (see Fig.~\ref{fig13}). Therefore,
the non-monotonic behaviour of the local density is responsible
for the non-monotonic behaviour of ${\cal{S}}$.

\section{Summary and discussion}

In the paper, we have studied analytically and numerically,
the spatial and $U$ dependence of short-range and
long-range ground-state correlation functions of electrons in the
Falicov-Kimball chain. We have chosen the electron-ion coupling $U>0$,
which amounts, in our case, to attraction between the electrons and ions.
The obtained results can be transformed to the case
of repulsion by means of hole-particle transformations \cite{grumac}.
We emphasize also that only nearest-neighbour, uncorrelated with ions,
hopping of electrons has been taken into account.
The properties of electron correlations in the case of extended or
correlated with ions hopping will be presented elsewhere.
The calculations have been carried out in different
ground states, specified by the sets $G$ of periodic ground-state ion
configurations and the corresponding stability intervals  $\Delta U_G$.
Typically, for a specified $G$ we considered an interval of $U$ having
only a non-void intersection with the stability interval of $G$, $\Delta U_G$.

Outside the stability intervals, the results obtained refer to
tight-binding electrons in periodic external potentials, given by $G$.

We mention that, due to the Jordan-Wigner transformation,
our results can also be interpreted in terms of spin-$1/2$ isotropic
$XY$ chains, in periodic transverse external magnetic fields.

It is well known that
the two-point correlation function $\la a^+_x a_{x+l} \ra_b$,
$l=0,1,\ldots$,
amounts to the one-body reduced density matrix in the ground state.
This quantity enables one to calculate ground-state averages of one-body
observables \cite{martin}, and is therefore of crucial importance for all
methods that describe many-body systems by means of single-particle
formalism, like  the density functional theory, for instance.
The spatial decay of the one-body reduced density matrix determines the
degree of locality of many quantities, which are relevant for describing
properties of metals and insulators. For this reason,
its asymptotic long-distance behaviour has been the subject of interest
for long time. Quite recently, the spatial decay of the one-body reduced
density matrix was investigated in some tight-binding models of metals and
insulators (see \cite{beigi}, \cite{taraskin1}, \cite{taraskin2},
and the references quoted there).
In \cite{taraskin1} the authors considered a model of an insulator
with two-bands separated by a gap, due to two bare electronic states,
with different bare energies, at each site, with equal in-band transfer
integrals and a weak interband hybridization. The asymptotic long-distance
behaviour of the single-particle density matrix has been studied,
in dimensions $D=1,2$, and $3$. The dependence of the correlation (decay)
length on the energy parameters of the model and the existence of the
power-law prefactor, of the form $l^{-D/2}$, has been demonstrated.
The results obtained in our paper can be
viewed as the results referring to models of many-band insulators, where the
bands separated by gaps arise due to a periodic spatial modulation of the bare
energies. In the case of two-bands (Section 4),
we have been able to calculate exactly the correlation functions,
and then extract, in particular, the asymptotic long-distance
behaviour of the single-particle density matrix.
In the considered here case of $D=1$,
we have found the $l^{-1/2}$ prefactor, as in \cite{taraskin1},
and the exact expression for the correlation length $\xi$,
given by (\ref{ksi}),
as a function of the periodic-potential strength $U$
(which amounts also to the gap width). For a weak potential,
the inverse correlation length $\xi^{-1} \approx U/2$,
while for a strong potential, $\xi^{-1} \approx 2\ln(U/2)$.

In \cite{macedo}, Mac\^{e}do et al calculated the correlation function
${\cal{S}}(l)$, given by (\ref{Splchessinfty}), as the $T \to 0$ limit of the
corresponding grand-canonical correlation function, that is, with
the ground-state symmetric average in definition (\ref{Sbx}) replaced by
the grand canonical average. The chemical potential was chosen in such a way
that the ground-state was characterized by the set $[1/2]_1$ of checkerboard
ion configurations. In the case of the one-dimensional system under consideration,
the grand canonical correlation functions should coincide with the canonical
ones. The results in \cite{macedo} have been obtained  using the method of
small-cluster exact diagonalization and extrapolation techniques to the
infinite chain, where the cluster size was limited to at most 10 sites.
To compare the values of ${\cal{S}}(l)$, obtained from the exact formula
(\ref{Splchessinfty}), and those displayed in figures of \cite{macedo},
one has to change the sign of $U$ in (\ref{Splchessinfty}), and to multiply
our ${\cal{S}}(l)$ by $3/4$. The results of exact calculations are shown in
Table~1, and they differ qualitatively from the data displayed in \cite{macedo}.

\begin{table}
\begin{center}
\caption[]{$3{\cal{S}}(l)/4$ }
\begin{tabular}[c]{|r|r|r|r|} \hline
$\ \ l\ \ $ & $\ \ U=0\ \ $    & $\ \ U=-4\ \ $    & $\ \ U=-8\ \    $ \\ \hline
$\ \ 0\ \ $ & $\ \ 0.375\ \ $  & $\ \ 0.688\ \ $  & $\ \ 0.729\ \  $ \\
$\ \ 1\ \ $ & $\ \ -0.236\ \ $ & $\ \ -0.658\ \ $ & $\ \ -0.719\ \ $ \\
$\ \ 2\ \ $ & $\ \ 0.188\ \ $  & $\ \ 0.630\ \ $  & $\ \ 0.710\ \  $ \\
$\ \ 3\ \ $ & $\ \ -0.196\ \ $ & $\ \ -0.632\ \ $ & $\ \ -0.710\ \ $ \\ \hline
\end{tabular}
\end{center}
\end{table}

The ground-state phase diagram of the Falicov-Kimball model is reach
in quantum phase transitions, that is the phase transitions where the role
of thermal fluctuations is played by quantum fluctuations \cite{sachdev1}.
These transitions are driven by such control parameters as $\rho_e$, $\rho_i$,
and $U$. They occur, when for some critical value of a control parameter,
the nature of the ground state changes. A detailed description of quantum
phase transitions in a many-body, interacting system, like the Falicov-Kimball
model, is a hard task (for recent results see \cite{messager2},
\cite{messager3}).
We would like to point out that the analytic results of Section 4, can be used
to describe a simple instance of a quantum phase transition, driven by $U$,
with the critical value $U_c=0$.
Consider an infinite chain, whose particle densities are set to $1/2$,
$\rho_i=\rho_e=1/2$.
Obviously, as $U=U_c=0$ there is no order in the ion subsystem,
whereas the electrons are distributed uniformly. There is no gap in the
electron energy spectrum and the truncated electron density-density
correlation function, $\langle n_xn_{x+l}\rangle - 1/4$,
exhibits a power-law decay $l^{-2}$
(with the oscillations owing to $\sin^2 (\pi l /2)$ imposed)
as $l\to\infty$, according to (\ref{nn0}).
On the other hand, for an arbitrary nonzero $U$,
the ion subsystem becomes checkerboard ordered
\cite{brasch, kennlieb}. The electrons follow the periodic distribution
of the ions, for instance the broken-symmetry average $\langle n_x\rangle_1$
is modulated with period $2$ (\ref{nx1}). A gap $\Delta=U$ appears
at the Fermi level, and correspondingly, the large-distance behaviour
of the broken-average truncated density-density correlation function,
$\langle n_xn_{x+l}\rangle_1$
$-\langle n_x\rangle_1\langle n_{x+l}\rangle_1$,
changes to an exponential one, of the form $l^{-1}\exp(-l/\xi)$
(see (\ref{a+a1evenasymp}), (\ref{a+a1oddasymp})),
with the correlation length $\xi \sim U^{-1}$, which diverges as $U \to 0$.

\vspace{5mm}

\noindent
{\bf Acknowledgements}
\\[1mm]
T. K. is grateful to the University of Wroc\l aw
for a kind hospitality during the completion of this manuscript.

\section{Appendix A}

Here we define the short- and long-range correlation functions
for the ion subsystem.
For any ion configuration $g \in G$, and its restriction to
a finite chain
whose number of sites, $L$, is a multiple of the period $q$ of
the ground-state configurations $g$,
\bea
q^{-1}\sum_{g\in G}g_x=\rho_i=L^{-1}\sum_{x=0}^{L-1} g_x .
\eea
The {\em short-range ion-ion correlation function} is
\bea
E_{b,\,x}(l)=q^{-1}\sum_{g\in G} \ld g_x-\rho_i \rd \ld
g_{x+l}-\rho_i \rd ,
\eea
which for $b=p$ is translation invariant.
The degree of order can be "measured" either by the
{\em {ion-order parameter}}, $I_{b,\,g}(k)$,
which for given $g \in G$ and $k=2\pi \rho_e = 2\pi p/ q$ is
\bea
I_{b,\,g}(k)=L^{-1}\sum_{x=0}^{L-1} \mbox{e}^{ikx} \ld g_x-\rho_i
\rd; \ \ \ \ \sum_g I_{b,\,g}(k)\equiv0 ,
\eea
or by the long-range correlation, $q^{-1}\sum_{g\in G}|I_{b,\,g}(k)|^2$,
\bea
q^{-1}\sum_{g\in G}|I_{b,\,g}(k)|^2=
L^{-2}\sum_{x,\,y=0}^{L-1}\mbox{e}^{ik(x-y)}
q^{-1}\sum_{g\in G} \left[ \ld g_x-\rho_i \rd \ld g_y-\rho_i \rd \right].
\eea
The function $k \rightarrow q^{-1}\sum_{g\in G}|I_{b,\,g}(k)|^2$ is the static
structure factor for the ions.
For periodic boundary conditions,
\bea
q^{-1}\sum_{g\in G}|I_{p,\,g}(k)|^2=
|I_{p,\,g}(k)|^2=
q^{-2}\sum_{x,\,y=0}^{q-1}\mbox{e}^{ik(x-y)}
\ld g_x-\rho_i \rd \ld g_y-\rho_i \rd > 0,
\eea
where the above inequality follows from the argument presented
in the Appendix of \cite{freefal}.
Thus, the square of the absolute value of the ion-order parameter
is a measure of the ion-ion long-range order.

\section{Appendix B}

For $\rho_e < 1/2$, the large $U$
asymptotic behaviour of $\lim_{L\to\infty}\tau_0(l)$ is smooth in $\rho_e$
and reads:
\bea
\mbox{for}\  U \to \infty, \ \
\lim_{L\to\infty}\tau_0(l) {\approx}\frac{1}{\pi\alpha}
\lb \frac{\sin(\pi\rho_e l)}{l}   \right. \non \left.
+\frac{1}{2\alpha^2} \lbr 2\frac{\sin(\pi\rho_e l)}{l}
-\frac{\sin(\pi\rho_e (l+2))}{l+2} - \frac{\sin(\pi\rho_e (l-2))}{l-2}\rbr \right. \non \left.
+\frac{3}{2\alpha^4} \lbr 3\frac{\sin(\pi\rho_e l)}{l}
-\frac{5}{2}\ld \frac{\sin(\pi\rho_e (l+2))}{l+2} + \frac{\sin(\pi\rho_e (l-2))}{l-2} \rd
\right. \right. \non \left.\left.
+\frac{\sin(\pi\rho_e (l+4))}{l+4} + \frac{\sin(\pi\rho_e (l-4))}{l-4}
\rbr \rb .
\eea
In particular,
\bea
\mbox{for}\ \ \  U \to \infty, \ \ \ \ \ \ \
\lim_{L\to\infty}\tau_0(0) {\approx} \rho_e/\alpha .
\eea
On the other hand, the small $U$
asymptotic behaviour of $\lim_{L\to\infty}\tau_0(l)$ is singular at $\rho_e=1/2$.
For $\rho_e < 1/2$, the following recursive formula holds:
\bea
\mbox{for}\ \ \ U \to 0, \ \ \ \ \ \ \
\lim_{L\to\infty}\tau_0(l) {\approx}\frac{\sin(\pi\rho_e (l-1))}{\pi(l-1)}
-\lim_{L\to\infty} \tau_0(l-2)\ .
\eea
Thus, knowing that
\bea
\mbox{for}\ \ \ U \to 0, \ \ \ \ \ \ \
\lim_{L\to\infty}\tau_0(0)
{\approx}\frac{1}{2\pi}\ln|\tan\lbr{\pi}(1+2\rho_e)/4 \rbr|\ ,
\eea
we can obtain the asymptotic formula for any even $l \geq 2$:
\bea
\lim_{L\to\infty}\tau_0(2n)\approx \frac{(-1)^n}{2\pi} \lb
\lbr 2\sum_{j=1}^n(-1)^j\frac{\sin(2j-1)\pi\rho_e}{2j-1}
 +\ln\tan\lbr{\pi}(1+2\rho_e)/4\rbr  \rbr \right. \non \left.
 +\frac{\alpha^2}{8}\lbr 4 \sum_{j=1}^{n-1}(-1)^j(n-j)(n-j+1)\frac{\sin(2j-1)\pi\rho_e}{2j-1}
\right.\right. \non \left.\left.
 +(2n^2-1/2)\ln\tan\lbr{\pi}(1+2\rho_e)/4\rbr
 -\frac{1}{2}\frac{\sin\pi\rho_e}{\cos^2\pi\rho_e}  \rbr + \ldots
\rb .
\eea
Then,  for $\rho_e = 1/2$ and $l=0,2$ we find:
\bea
\mbox{for}\ \ \ U \to 0, \ \ \ \ \ \ \
\lim_{L\to\infty}\tau_0(0) {\approx}
\frac{1}{2 \pi} ( a_0 + 2\ln2 - \ln \alpha ) ,
\eea
\bea
\mbox{for}\ \ \ U \to 0, \ \ \ \ \ \ \
\lim_{L\to\infty}\tau_0(2) {\approx}
\frac{1}{2 \pi} (2 - a_0 - 2\ln2 + \ln \alpha) ,
\eea
with $a_0$ given by
\bea
a_0=\frac{1}{2}\sum_{n=1}^{\infty}\frac{(2n-1)!!}{n(2n)!!}\approx 0.687 .
\eea

\section{Appendix C}
The relation between the integral in (\ref{tau0eveninf}) and the hypergeometric
function can be obtained as follows:
\bea
\label{passage}
\lim_{L \to \infty} \tau_0(2s)=
\frac{\kappa}{4\pi}\int\limits_0^{\pi}
\frac{\cos(2sk)}{\sqrt{1-\kappa^2\sin^2k}}\mbox{d}k
\non
=\frac{\kappa}{4\pi}\int\limits_0^{\pi}
\mbox{d}k\cos(2sk)\sum_{j=0}^{\infty}\frac{(2j-1)!!}{(2j)!!}\kappa^{2j}\sin^{2j}k
\non
=\frac{\kappa}{4\pi}\sum_{j=0}^{\infty}
\frac{(2j-1)!!}{(2j)!!}\kappa^{2j}\int\limits_0^{\pi}\mbox{d}k\cos(2sk)\sin^{2j}k
\non
=\frac{\kappa}{4\pi}\sum_{j=s}^{\infty}
\kappa^{2j}\frac{(2j-1)!!}{(2j)!!}(-1)^s\frac{\pi}{2^{2j}}\frac{(2j)!}{(j+s)!(j-s)!}
\non=(-1)^s\frac{\kappa}{4}\sum_{j=s}^{\infty}
\frac{(2j-1)!!}{(2j)!!}\,\frac{(2j)!}{(j+s)!(j-s)!}\,\frac{\kappa^{2j}}{2^{2j}}
\non=(-1)^s\frac{\kappa}{4}\sum_{j=s}^{\infty}
\frac{[(2j-1)!!]^2}{(j+s)!(j-s)!}\,\frac{\kappa^{2j}}{2^{2j}}
\non=(-1)^s\frac{\kappa}{4}\sum_{j=0}^{\infty}
\frac{[(2s+2j-1)!!]^2}{(2s+j)!}\,\frac{1}{j!}\,\frac{\kappa^{2(s+j)}}{2^{2(s+j)}}
\non
=(-1)^s
\frac{\kappa^{2s+1}}{4}\sum_{j=0}^{\infty}
\frac{\Gamma^2(s+\frac{1}{2}+j)}{\Gamma(2s+1+j)}\,\frac{\kappa^{2j}}{j!}
\non=(-1)^s\frac{\kappa^{2s+1}}{4\pi}\,
\frac{\Gamma^2(s+\frac{1}{2})}{\Gamma(2s+1)}F(s+\frac{1}{2},s+\frac{1}{2},2s+1;\kappa^2)
\non
=(-1)^s\frac{\kappa^{2s+1}}{2^{3s+2}}\,
\frac{(2s-1)!!}{s!}F(s+\frac{1}{2},s+\frac{1}{2},2s+1;\kappa^2) .
\eea

The Laplace asymptotic formula for integrals reads \cite{bo}:
\bea
\label{laplace}
\int\limits_0^1f(t)\mbox{e}^{sS(t)}\mbox{d}t
\stackrel{s\to\infty}{\approx}
\sqrt{-\frac{2\pi}{s S''(t_0)}}f(t_0)\mbox{e}^{sS(t_0)} ,
\eea
where
\bea
\label{max}
S(t_0)=\max_t S(t) .
\eea
In the case of the integral in the last line of (\ref{tau0evenLap}),
\bea
\label{fS}
f(t)\equiv\frac{1}{\sqrt{t(1-t)(1-\kappa^2t)}}, \ \ \ \ \
S(t)\equiv\ln\frac{t(1-t)}{1-\kappa^2t} ,
\eea
with
\bea
\label{t0}
t_0=\frac{1-\sqrt{1-\kappa^2}}{\kappa^2},
\eea
and
\bea
\label{att0}
S(t_0)=\ln(t_0^2)=\ln\lbr
\frac{1-\sqrt{1-\kappa^2}}{\kappa^2}\rbr^2, \ \ \ \
f(t_0)=\frac{\kappa^2}{(1-\sqrt{1-\kappa^2})\sqrt{1-\kappa^2}}, \
\ \  \non
S''(t_0)=-\frac{2\kappa^4}{\sqrt{1-\kappa^2}(1-\sqrt{1-\kappa^2})^2} .
\eea
Consequently, the asymptotic form of the considered integral is
\bea
\label{asintegral}
\sqrt{\frac{\pi}{s}}\,\frac{1}{\sqrt[4]{1-\kappa^2}}\,
\lbr \frac{1-\sqrt{1-\kappa^2}}{\kappa^2} \rbr^{2s}=
\sqrt{\frac{\pi}{s}}\,\sqrt[4]{\frac{1-(\frac{U}{4})^2}{(\frac{U}{4})^2}}\,
\mbox{e}^{-s/\xi} ,
\eea
with $\xi$ given by (\ref{ksi}).

\clearpage

\begin{figure*}[t]
\centerline{\includegraphics[clip,width=8cm]{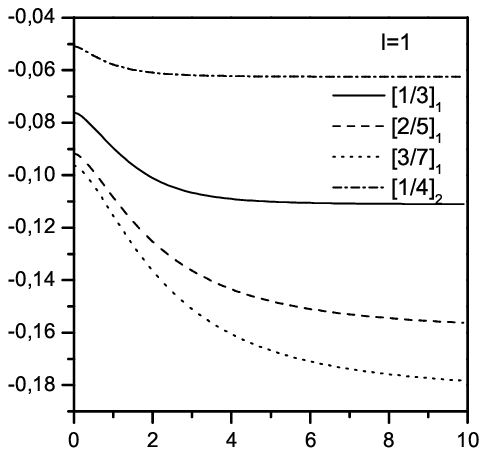}
\hfill
\includegraphics[clip,width=8cm]{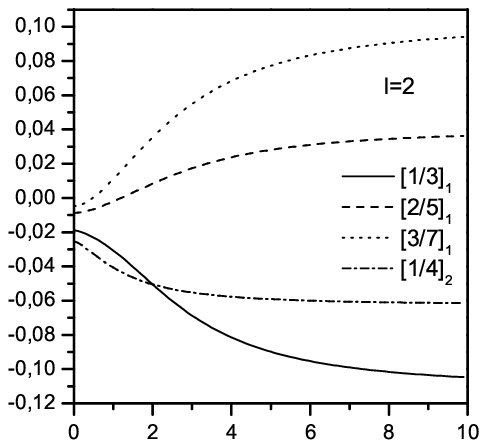}}
\vspace{-15mm}
\centerline{\includegraphics[clip,width=8cm]{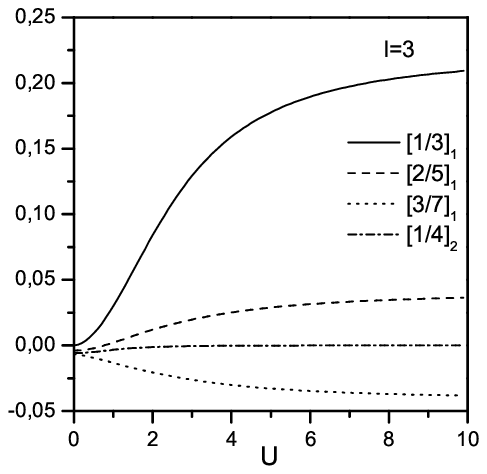}
\hfill
\includegraphics[clip,width=8cm]{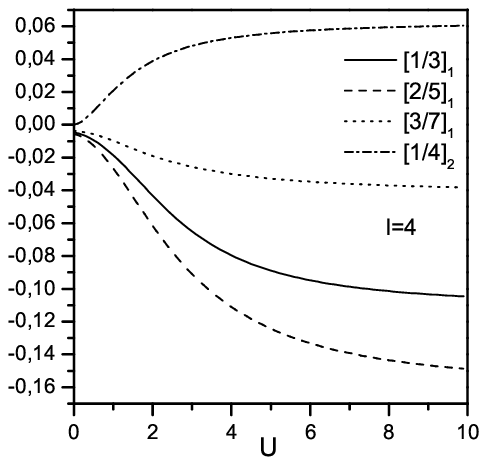}}
\caption[]
{Infinite chain limit of ${\cal{L}}_{f,x}(l)$, given by (\ref{Lbx}),
versus $U$, for different distances $l$, and different sets $G$.
}
\label{fig9}
\end{figure*}

\begin{figure*}[t]
\centerline{\includegraphics[clip,width=8cm]{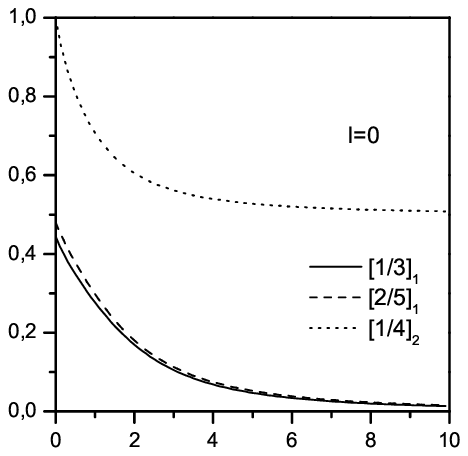}
\hfill
\includegraphics[clip,width=8cm]{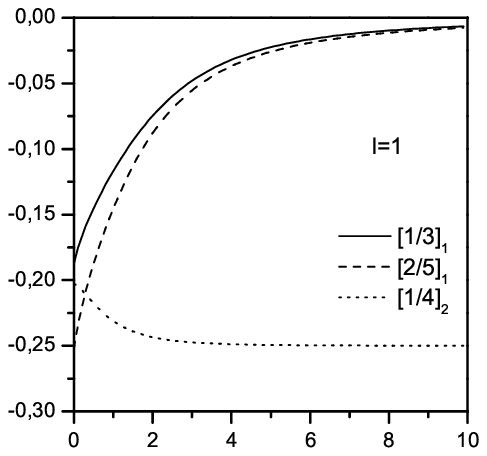}}
\vspace{-15mm}
\centerline{\includegraphics[clip,width=8cm]{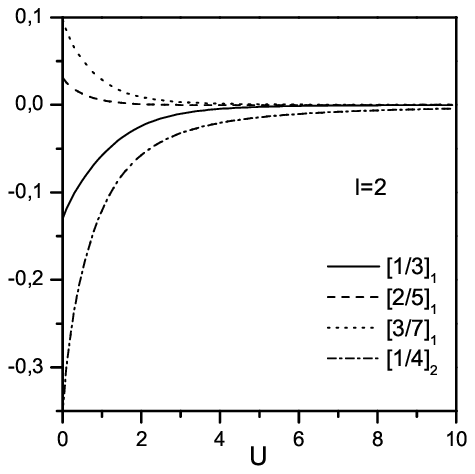}
\hfill
\includegraphics[clip,width=8cm]{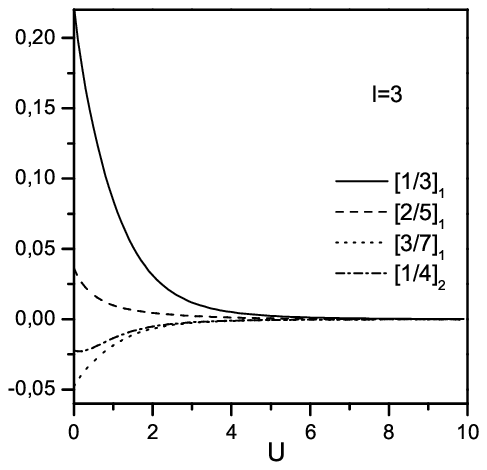}}
\caption[]
{Infinite chain limit of ${\cal{S}}_{f,x}(l)$, given by (\ref{Sbx}),
versus $U$, for different distances $l$, and different sets $G$.
Whenever a plot for $G=[3/7]_1$ is missing, it is indistinguishable
from the plot for $G=[2/5]_1$, in the scale of the figure.
}
\label{fig10}
\end{figure*}

\begin{figure*}[t]
\centerline{\includegraphics[clip,width=8cm]{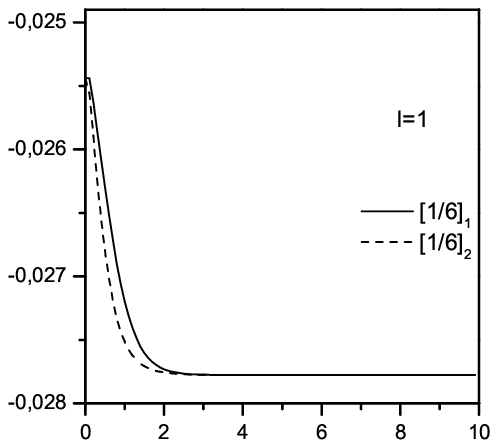}
\hfill
\includegraphics[clip,width=8cm]{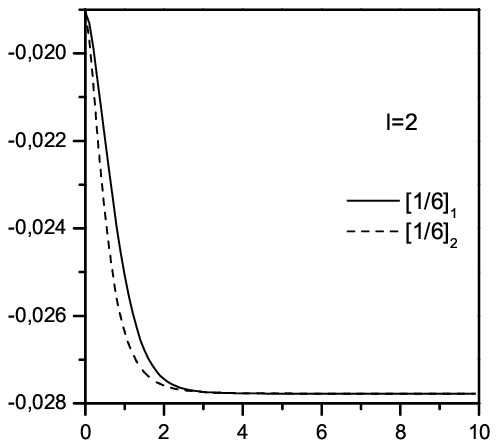}}
\vspace{-15mm}
\centerline{\includegraphics[clip,width=8cm]{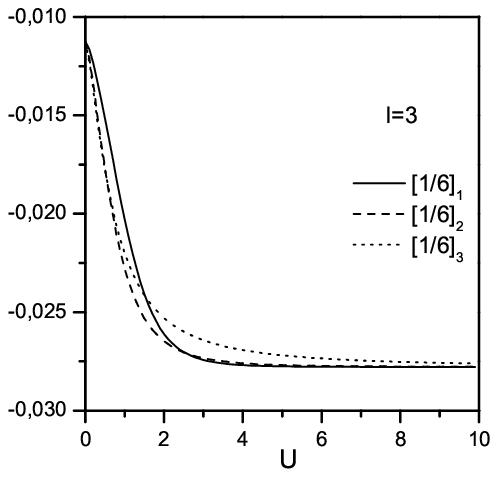}
\hfill
\includegraphics[clip,width=8cm]{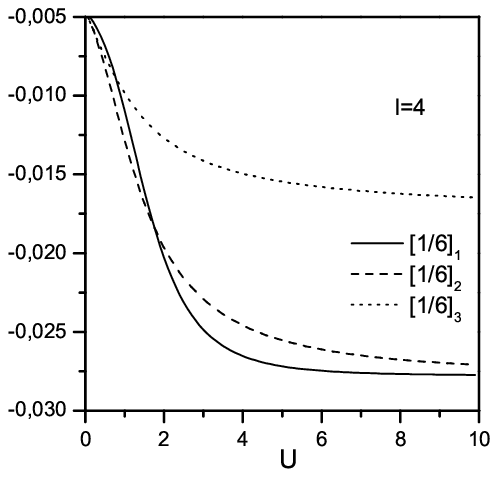}}
\caption[]
{Infinite chain limit of ${\cal{L}}_{f,x}(l)$, given by (\ref{Lbx}),
versus $U$, for different distances $l$, and different sets $G$.
Whenever a plot for $G=[1/6]_3$ is missing, it is indistinguishable
from the plot for $G=[1/6]_2$, in the scale of the figure.
}
\label{fig11}
\end{figure*}

\begin{figure*}[t]
\centerline{\includegraphics[clip,width=8cm]{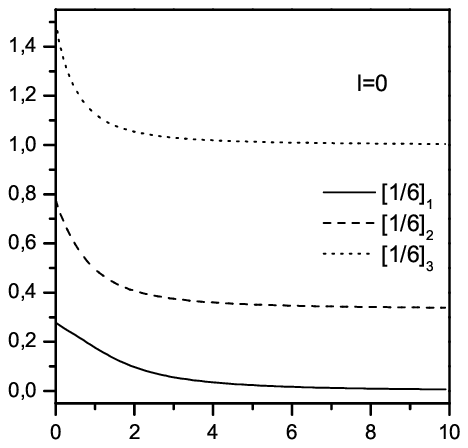}
\hfill
\includegraphics[clip,width=8cm]{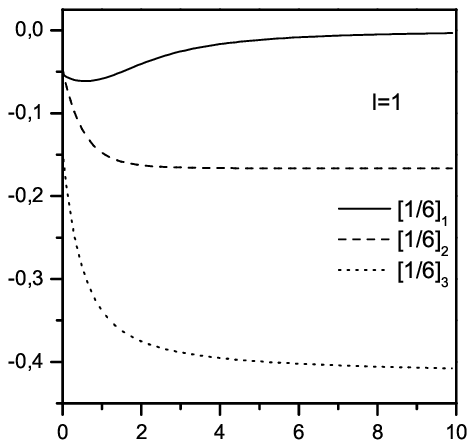}}
\vspace{-15mm}
\centerline{\includegraphics[clip,width=8cm]{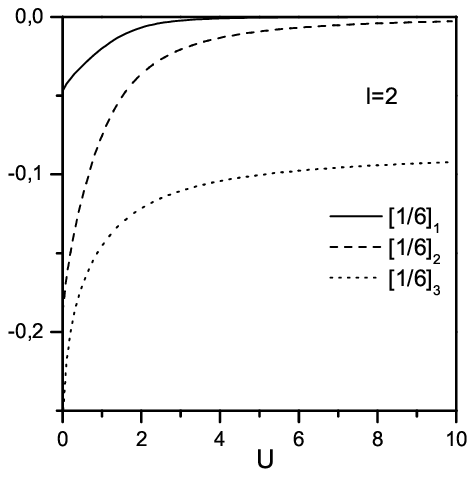}
\hfill
\includegraphics[clip,width=8cm]{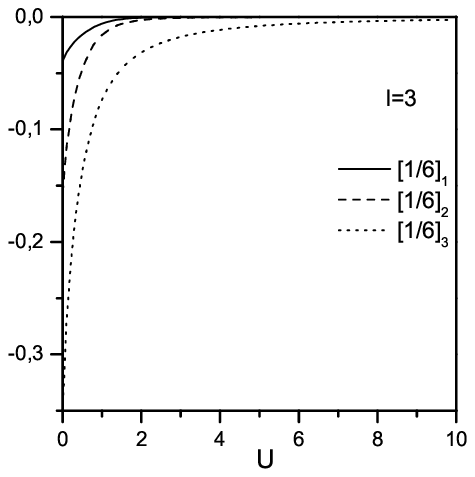}}
\caption[]
{Infinite chain limit of ${\cal{S}}(l)$, given by (\ref{Sbx}),
versus $U$, for different distances $l$, and different sets $G$.
}
\label{fig12}
\end{figure*}

\begin{figure}[t]
\centerline{\includegraphics[clip,width=8cm]{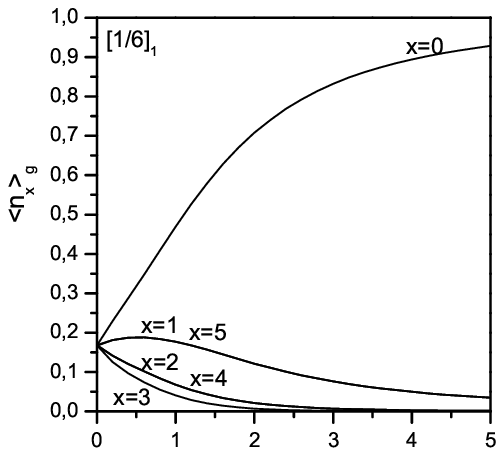}}
\vspace{-13mm}
\centerline{\includegraphics[clip,width=8cm]{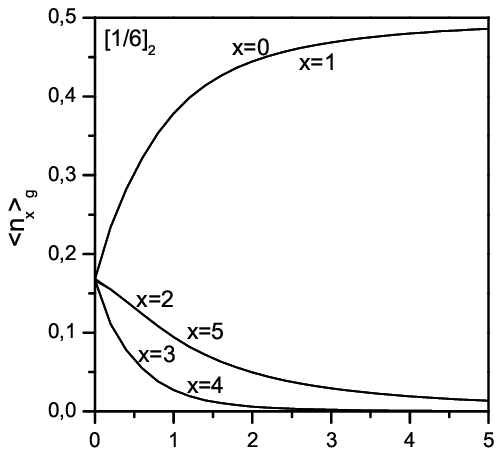}}
\vspace{-13mm}
\centerline{\includegraphics[clip,width=8cm]{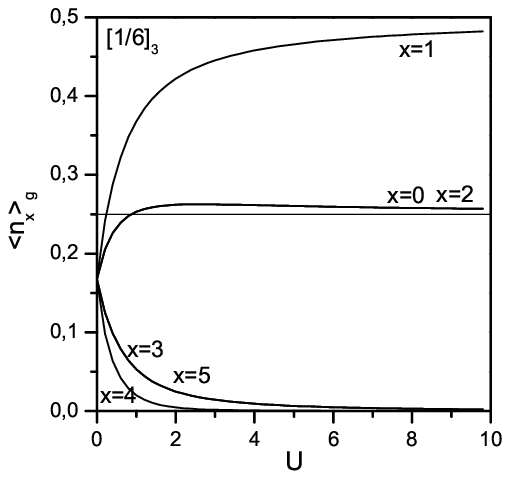}}
\caption[]
{The local electron density in a fixed ion configuration $g$,
versus position $x$, as a function of $U$.
The ion configuration
$g=100000$ for $G=[1/6]_1$, $g=110000$ for $G=[1/6]_2$,
and $g=111000$ for $G=[1/6]_3$.}
\label{fig13}
\end{figure}

\begin{figure}[t]
\centerline{\includegraphics[clip,width=8cm]{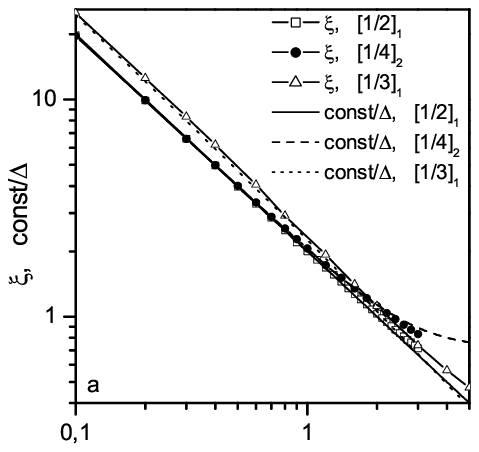}}
\vspace{-1cm}
\centerline{\includegraphics[clip,width=8cm]{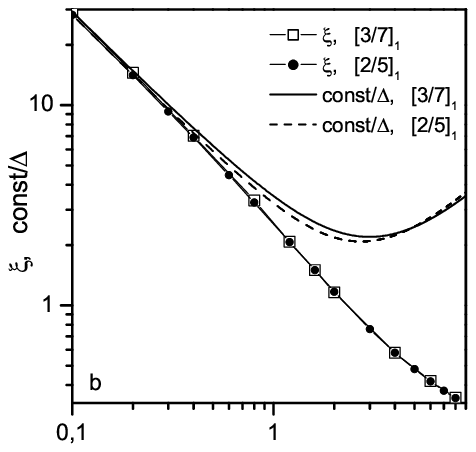}}
\vspace{-1cm}
\centerline{\includegraphics[clip,width=8cm]{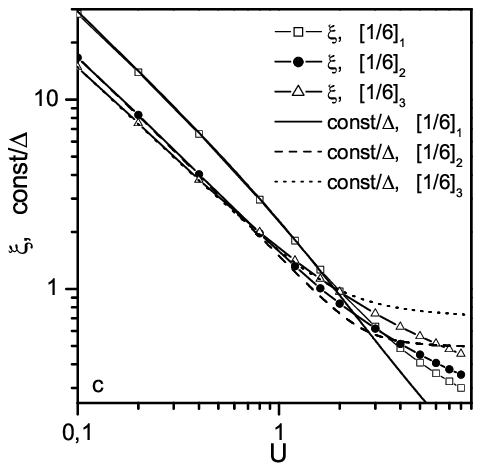}}
\caption[]
{Numerically calculated correlation length $\xi$ versus $U$,
and numerically calculated $const \Delta^{-1}$,
with $const$ adjusted so that $\xi \approx const \Delta^{-1}$,
for different sets $G$.
The $\xi$ has been obtained from approximating the large-distance
behaviour of $|\la a^+_x a_{x+l} \ra_b|^2$ by $const \exp(-l/\xi)$.
}
\label{fig14}
\end{figure}

\begin{figure}[t]
\centerline{\includegraphics[clip,width=8cm]{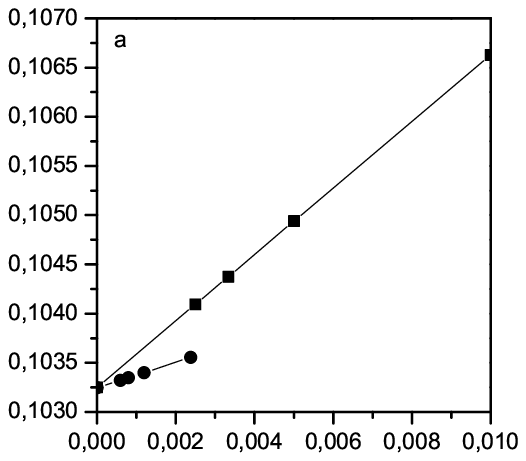}}
\vspace{-12mm}
\centerline{\includegraphics[clip,width=8cm]{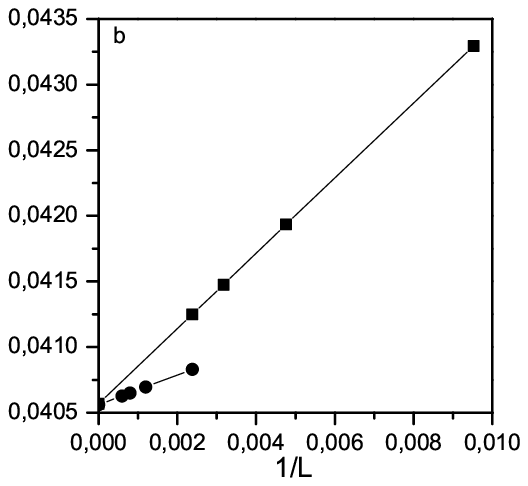}}
\caption[]
{${\cal{P}}_f(2\pi \rho_e)$, given by (\ref{Pb}), and
${\cal{P}}'_f(2\pi \rho_e)$, given by (\ref{P'b}), versus $L^{-1}$,
for $U=2$, and $G=[1/2]_1$ (a), $G=[3/7]_1$ (b).
Filled squares: ${\cal{P}}_f(2\pi \rho_e)$,
filled circles: ${\cal{P}}'_f(2\pi \rho_e)$.}
\label{fig15}
\end{figure}

\begin{figure}[t]
\centerline{\includegraphics[clip,width=8cm]{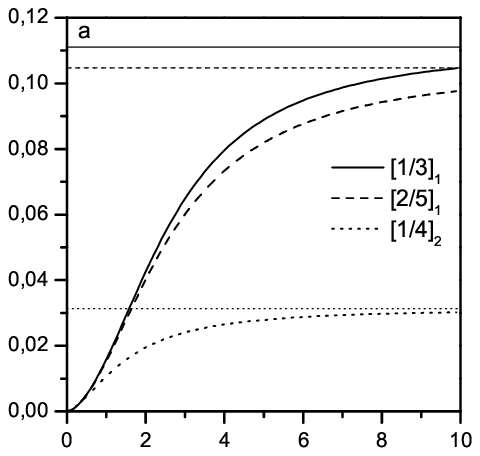}}
\vspace{-13mm}
\centerline{\includegraphics[clip,width=8cm]{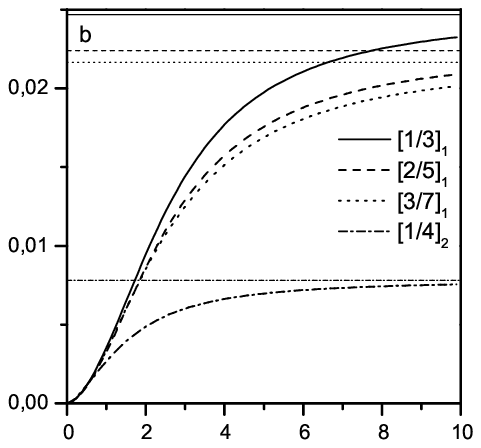}}
\vspace{-15mm}
\centerline{\includegraphics[clip,width=8cm]{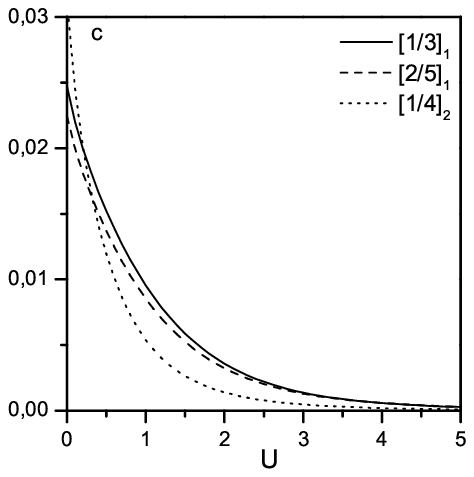}}
\vspace{-10mm}
\caption[]
{(a) ${\cal{P}}(2\pi \rho_e)$, given by (\ref{P}),
versus $U$, for different sets $G$,
(b) ${\cal{L}}$, given by (\ref{L}), versus $U$, for different sets $G$,
(c) ${\cal{S}}$, given by (\ref{S}), versus $U$, for different sets $G$.
In (a) and (c) the plot corresponding to $G=[3/7]_1$ is missing, since it is
indistinguishable from the plot for $G=[2/5]_1$.
}
\label{fig16}
\end{figure}

\begin{figure*}[t]
\centerline{\includegraphics[clip,width=8cm]{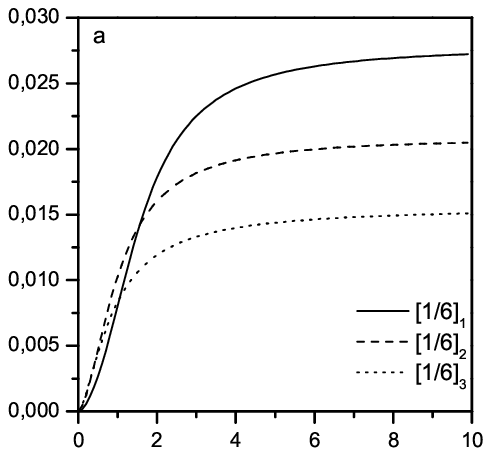}}
\vspace{-15mm}
\centerline{\includegraphics[clip,width=8cm]{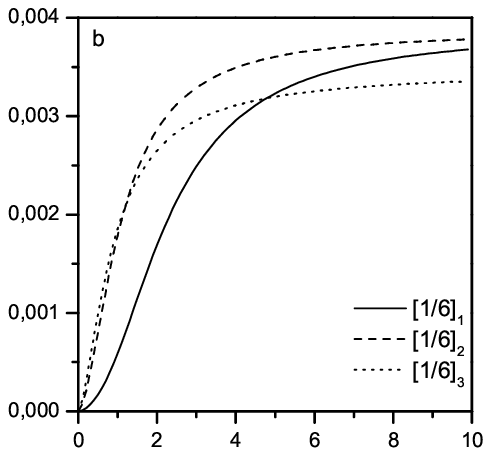}}
\vspace{-15mm}
\centerline{\includegraphics[clip,width=8cm]{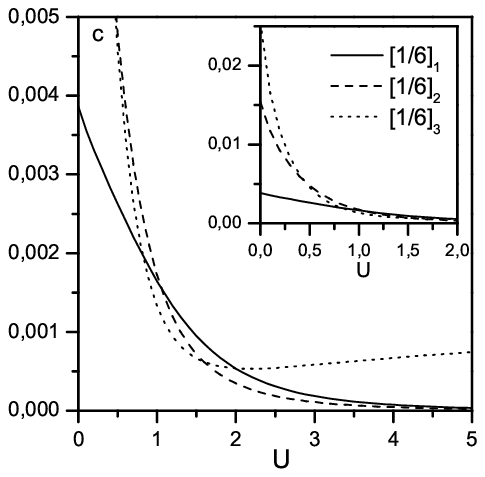}}
\vspace{-10mm}
\caption[]
{(a) ${\cal{P}}(2\pi \rho_e)$, given by (\ref{P}),
versus $U$, for different sets $G$,
(b) ${\cal{L}}$, given by (\ref{L}), versus $U$, for different sets $G$,
(c) ${\cal{S}}$, given by (\ref{S}), versus $U$, for different sets $G$.
}
\label{fig17}
\end{figure*}

\clearpage


\end{document}